\documentclass[english,a4paper,12pt]{article}
\usepackage[utf8]{inputenc}
\usepackage[T1]{fontenc}
\usepackage{titling}

\usepackage{amsmath, amssymb}
\usepackage{pgf}
\usepackage{amsfonts}
\usepackage{dsfont}
\usepackage{mathrsfs}
\usepackage{mathabx}
\usepackage{stmaryrd}
\usepackage{makeidx}
\usepackage{amsbsy}
\usepackage{amsthm}
\usepackage{color}
\usepackage{fullpage}
\usepackage{enumitem}
\usepackage[vcentermath]{youngtab}
\usepackage{marginnote} 
\usepackage{mdframed}	
\newmdenv[topline=false, bottomline=false, skipabove=\topsep, skipbelow=\topsep]{siderules}
\usepackage{tikz}
\usetikzlibrary{snakes}
\usetikzlibrary{decorations.text,calc,arrows.meta}
\usepackage[vcentermath]{youngtab}
\usepackage{tkz-euclide}
\usepackage{pgfplots}
\usetikzlibrary{decorations.shapes}
\usetikzlibrary{positioning}
\usetikzlibrary{patterns}
\usetikzlibrary{shapes.geometric}
\usetikzlibrary{decorations.pathmorphing}
\usetikzlibrary{arrows.meta}
\tikzset{snake it/.style={decorate, decoration=snake}}
\usetikzlibrary{arrows,shapes,positioning}
\usetikzlibrary{decorations.markings}
\tikzstyle arrowstyle=[scale=1]
\tikzstyle directed=[postaction={decorate,decoration={markings,mark=at position .65 with {\arrow[arrowstyle]{stealth}}}}]
\tikzstyle reverse directed=[postaction={decorate,decoration={markings,mark=at position .65 with {\arrowreversed[arrowstyle]{stealth};}}}]

\def\sH{\mathscr{H}}

\def\cA{{\mathfrak A}}
\def\cB{{\mathfrak B}}

\def\cN{{\mathfrak N}}

\newcommand{\A}{\mathfrak{A}}

\def\cA{{\ca A}}
\def\cB{{\ca B}}

\def\cE{{\ca E}}

\def\cN{{\ca N}}

\def\cR{{\ca R}}

\newcommand{\ca}[1]{{\cal #1}}
\newcommand{\ben}{\begin{equation}}
\newcommand{\een}{\end{equation}}
\def\bena{\begin{eqnarray}}
\def\eena{\end{eqnarray}}

\def\1{{\mathds{1}}}

\newcommand{\tr}{\operatorname{Tr}}

\renewcommand{\log}{\operatorname{ln}}

\renewcommand{\epsilon}{\varepsilon}

\newcommand{\CC}{\mathbb{C}}

\begin{document}
\title{
GGI Lectures on Entropy, Operator Algebras and Black Holes
}

	\author{Stefan Hollands$^{1}$\thanks{\tt stefan.hollands@uni-leipzig.de}\\
	{\it $^1$ Universit\" at Leipzig, ITP and MPI-MiS Leipzig}
	}

\date{\today}
	
\maketitle

\begin{abstract}
These are extended notes for my lectures at the workshop ``Reconstructing the Gravitational Hologram'' held at Galileo-Galilei Institute, Florence, in June 2022. 
\end{abstract}


\section{Hawking, Unruh, Bisognano-Wichmann}

The aim of these notes is to outline some interesting recent and not so recent developments at the interface between high energy physics and gravity, involving directly or indirectly ideas about 
entropy and quantum information. Emphasis is in particular on the connection to operator algebras. I am aware that operator algebras are fairly unfamiliar territory for 
most practitioners in high energy physics and quantum information, so my exposition will be quite informal for the most part, drawing attention to some methods and notions that I believe can be useful 
also for someone with only a casual interest in the technicalities of the subject. Before I begin let me emphasize that notions related to entropy have been used by mathematicians working in 
operator algebras almost from the beginning. These lectures by no means represent an exhaustive expos\'e of all these highly non-trivial connections. 
As these were covered by other excellent lectures at the workshop and there is an abundant literature, I also will not cover path integral approaches to entropies in QFT and quantum gravity, which
can give very elegant and suggestive 
replica-trick representation of many quantities related to entanglement entropies. 

\medskip

Let me start, then, with the quintessential picture of Special Relativity ($x^\pm = x^1 \pm x^0$)
representing a Lorentz boost,
\ben
\begin{split}
x^0(\tau) =& \cosh (a \tau) x^0 + \sinh(a \tau) x^1\\
x^1(\tau) =& \sinh (a \tau) x^0 + \cosh(a \tau) x^1.
\end{split}
\een

\begin{figure}[h!]
\centering
\begin{tikzpicture}[scale=.5]
\draw[->, thick, dashed] (-2,-3) -- (6,5);
\draw[->,thick, dashed] (6,-3) -- (-2,5);
\draw (-1.7,5) node[anchor=west]{$x_-$};
\draw (5.7,5) node[anchor=east]{$x_+$};
\draw[<-,thick] (-2.5,-2.5) .. controls (1,1)  .. (-2.5,4.5);
\draw[->,thick] (6.5,-2.5) .. controls (3,1)  .. (6.5,4.5);
\end{tikzpicture}
\end{figure}

I am using units where the speed of light is $c=1$, so $a$ can be thought of as an acceleration parameter. Let 
\ben
\beta = \frac{2\pi}{a}
\een
By the addition theorem for $\sinh$ and $\cosh$:
\ben
\begin{split}
x^0(\tau + i\beta) =& x^0(\tau)\\
x^1(\tau +i\beta) =& x^1(\tau).
\end{split}
\een
This suggests that vacuum state correlation functions of a quantum field $\phi$ should be periodic when expressed in terms of $\tau$,  provided that their analytic 
continuation in $\tau$ makes sense. It turns out that this is the case precisely if we restrict the points in the correlation function to all 
lie in one of the wedges. The point is that in a wedge, the boost time has a definite direction; it is future directed
in the right wedge $\pm x^\pm > 0$. You are invited to check this in the following
exercise.

\medskip
\noindent
{\bf Exercise 1:}
Check that the imaginary part of the vector $(x^0(\tau + i\sigma), x^1(\tau + i\sigma), z^A)$ is in the future lightcone for $0<\sigma<\beta$ if $(x^0, x^1, z^A)$ is in the right wedge. Here $z^A$ are 
the coordinates along the edge of the wedge.
In a quantum field theory (QFT) with vacuum vector $|\Omega\rangle$, it is usually assumed that the generator $P^\mu$ of spacetime translations has 
spectrum in the forward light cone and $P^\mu |\Omega\rangle = 0$, meaning that the eigenvalues $p^\mu$ of $P^\mu$ satisfy $p^\mu p_\mu \le 0$ and $p^0 \ge 0$. 
By considering a suitable matrix element of $e^{ia P}$ and using 
the Heisenberg equation $e^{iaP} \phi(x) e^{-iaP} = \phi(x+a)$, show that the correlation function $\langle \Omega| \phi(x) \phi(x')|  \Omega \rangle$ has an analytic continuation $x' \to x' + iy'$ and $x \to x +iy$ provided that 
$y'$ is in the future- and $y$ is in the past lightcone. 

\medskip
\noindent
Thus in the right wedge the correlation function of a Minkowski QFT has the following periodicity in imaginary boost time $\tau$:
\begin{figure}[h!]
\centering
\begin{tikzpicture}[scale=.5]
\draw[->, thick, dashed] (-2,-3) -- (6,5);
\draw[->,thick, dashed] (6,-3) -- (-2,5);
\filldraw[color=black, fill=lightgray, thin] (6,-3) -- (10,1) -- (6,5) -- (2,1) -- (6,-3);
\draw (-1.7,5) node[anchor=west]{$x_-$};
\draw (5.7,5) node[anchor=east]{$x_+$};
\draw[->,thick] (6,-2.5) .. controls (2.5,1)  .. (6,4.5);
\draw (5,1) node[anchor=west]{$W$};
\draw (-1,1) node[anchor=east]{$W'$};
\end{tikzpicture}
\end{figure}
\ben
\langle \Omega | \phi(\tau) \phi(\tau' + i\beta - i0) | \Omega \rangle = \langle \Omega | \phi(\tau'-i0) \phi(\tau) | \Omega \rangle.
\een
The order of the operators has changed because we approach the real $x^0$-axis from above resp. below on the left resp. right sides of this equation -- you can 
check this by being careful about the ``$i0$''-prescriptions in these correlators.  
I have omitted the coordinates other than the boost time $\tau$ for easier readability. These other coordinates are to be held fixed.

It is important to note that we would get the same periodicity -- also called the KMS condition -- if we had for $x,x'$ in the right wedge $W$
\ben
{}_{W'W}\langle \Omega | \phi(x) \phi(x') | \Omega \rangle_{W'W} \, \propto \, \tr_W \phi(x) \phi(x') e^{-\beta K} 
\een
where $K$ is the generator of boosts in the right wedge $W$, and where $W'$ is the causal complement of the right wedge, i.e. the left wedge. 
\begin{figure}[h!]
\label{fig:1}
\centering
\begin{tikzpicture}[scale=.5]
\draw[->, thick, dashed] (-2,-3) -- (6,5);
\draw[->,thick, dashed] (6,-3) -- (-2,5);
\filldraw[color=black, fill=lightgray, thin] (6,-3) -- (10,1) -- (6,5) -- (2,1) -- (6,-3);
\draw (-1.7,5) node[anchor=west]{$x_-$};
\draw (5.7,5) node[anchor=east]{$x_+$};
\draw[thick, red] (2,1) -- (10,1);
\draw (3.7,0.5) node[anchor=west]{\textcolor{red}{$x^0=0,x^1>0$}};
\end{tikzpicture}
\end{figure}
Here, I have formally assumed that 
\ben
{\mathscr H} = {\mathscr H}_{W'} \otimes {\mathscr H}_{W}, \quad \text{(formal)}
\een
because the wedge $W$ and the opposite wedge $W'$ are causally disjoint, and to emphasize that this is a vector of the full Hilbert space,  
$|\Omega\rangle_{W'W}$ is indexed by the union $W'W$ of both wedges, the causal closure of which is the full spacetime. 
Also, formally, the generator of boosts in the right wedge ($x^1>0$) is:
\ben
K = a\int_{\textcolor{red}{\Sigma}} (x^1 \Theta_{0}{}^0 + x^0 \Theta_{1}{}^0) d^3 x 
\een
where $\Theta_{\mu}{}^{\nu}$ is the stress energy tensor of the QFT.
$\Sigma$ is any Cauchy surface for the wedge $W$, e.g.,, $\Sigma = \{ (0,x^1,x^2,x^3) \ : \ x^1 > 0\}$ indicated by the red line in figure \ref{fig:1}.
The conclusion of this reasoning is:

\medskip
\noindent
{\bf Unruh-Hawking:}
To an observer moving with constant acceleration $a$, the vacuum state looks thermal with Unruh-Hawking temperature
\ben
T = \frac{a}{2\pi}.
\een
The thermal quanta can be harvested by coupling the quantum field to a so-called Unruh-DeWitt detector moving on the uniformly accelerated 
trajectory. If the detector represents a BBQ party on a rocket, then the sausages get cooked provided $a$ is sufficiently high. You do not want to 
to be a guest at this party of course because you would get cooked, too. 

\medskip
\noindent
The above argument (with an important technical caveat regarding the status of the ``density matrix'' $e^{-\beta K}$ and the factorization 
of $\sH$ into $\sH_W \otimes \sH_{W'}$) 
as presented is essentially due to Bisognano and Wichmann \cite{bw}, who however did not themselves make the connection with the Unruh effect. 
Instead, their motivation was to establish a technical property of the observable algebras $\cA_W$ associated with the right wedge called ``Haag duality'' 
which I will explain later.

Almost literally the same reasoning applies to a Schwarzschild black hole
\ben
ds^2 = -f d\tau^2 + f^{-1}dr^2 + r^2 d\Omega^2, \quad f=1-\frac{2MG}{r}.
\een
\begin{figure}[h!]
\centering
\begin{tikzpicture}[scale=.5]
\draw[snake] (-2,5) -- (6,5);
\draw[snake] (6,-3) -- (-2,-3);
\filldraw[color=black, fill=lightgray, thin] (6,-3) -- (10,1) -- (6,5) -- (2,1) -- (6,-3);
\filldraw[color=black, fill=white, thin] (-2,-3) -- (-6,1) -- (-2,5) -- (2,1) -- (-2,-3);
\draw[ thick, dashed] (-2,-3) -- (6,5);
\draw[thick, dashed] (6,-3) -- (-2,5);
\draw (3.7,3) node[anchor=east]{$H^+$};
\draw (3.7,-1) node[anchor=east]{$H^-$};
\draw (8.4,3) node[anchor=west]{$I^+$};
\draw (8.4,-1) node[anchor=west]{$I^-$};
\draw[->,thick] (5.7,-2.3) .. controls (2.5,1)  .. (5.7,4.3);
\draw[->,thick] (6.3,-2.3) .. controls (9.5,1)  .. (6.3,4.3);
\end{tikzpicture}
\end{figure}
Now the right wedge is the exterior of the black hole $r>r_0=2MG$ (outside the Schwarzschild radius $r_0$). 
The boosts correspond to shifts in $\tau$. The full picture is best constructed using Kruskal-Szekeres coordinates
which are analogous to $x^\pm$. 

An observer just hovering along the event horizon $H^+$ (or $H^-$) given by $r=r_0$ must be accelerated with 
\ben
a = \frac{MG}{r_0^2} 
\een
which is an expression of Newton's law. 
\begin{figure}[h!]
\centering
\begin{tikzpicture}[scale=.5]
\draw[snake] (-2,5) -- (6,5);
\draw[snake] (6,-3) -- (-2,-3);
\filldraw[color=black, fill=lightgray, thin] (6,-3) -- (10,1) -- (6,5) -- (2,1) -- (6,-3);
\filldraw[color=black, fill=white, thin] (-2,-3) -- (-6,1) -- (-2,5) -- (2,1) -- (-2,-3);
\draw[ thick, dashed] (-2,-3) -- (6,5);
\draw[thick, dashed] (6,-3) -- (-2,5);
\draw (3.7,3) node[anchor=east]{$H^+$};
\draw (3.7,-1) node[anchor=east]{$H^-$};
\draw (8.4,3) node[anchor=west]{$I^+$};
\draw (8.4,-1) node[anchor=west]{$I^-$};
\draw[->,thick] (2,1)  -- (5,4);
\draw[->,thick] (5,-2)  -- (2,1);
\end{tikzpicture}
\end{figure}
By contrast to the Unruh effect where $a$ is a property of the observer, $a$ in the present context (also called the ``surface gravity'') is fixed by requiring that the $\tau$-coordinate be normalized so that 
\ben
ds^2 \sim -d\tau^2 + dr^2 + r^2 d\Omega^2 \quad \text{as $r \to \infty$,}
\een
so $\tau$ becomes the usual inertial time $x^0$ near infinity, due to gravitational redshift. Thus, unlike in the Unruh effect, in the black hole case, $a$ can be thought of as an intrinsic property of the spacetime. Nevertheless, either in the black hole case or the wedge, 
we can formally summarize the situation by saying that the reduced density matrix 
for the right wedge $W$ is 
\ben
 \omega_{W} = \tr_{W'} |\Omega\rangle \langle \Omega | \, \propto \, e^{-\beta K},  
\een
with $\beta = 2\pi/a$ the inverse Hawking-temperature and with trace taken over the opposite wedge $W'$. According to the 2nd law of black hole thermodynamics, the 
entropy of the black hole alone is given by 
\ben
S = \frac{{\rm horizon \, area}}{4G} \qquad 
\een
in units where $k_B = \hbar = c = 1$ with $4\pi r_0^2$ the area of the event horizon. This begs the question whether the entropy associated with $\omega_W$ (for all quantum fields including e.g., gravitions) could perhaps be equal to the area of the event horizon. This naive expectation turns out to be false as stated but is, as I hope to convince you, nevertheless a fruitful starting point for some considerations. 

\section{Entropy and relative entropy}

The surprise we experience at seeing an event is greater if we thought that it was less likely. Thinking about our surprise at seeing two independent events such as winning the 
lottery twice, it is plausible that our surprise should in this case be additive, not multiplicative. Thus, the surprise at seeing an event whose probability we think is $p$ should be 
proportional to $\log \frac{1}{p}$. Hence the average surprise experienced if events $x \in X$ are distributed according to a probability distribution $\underline{p} = (p_1, \dots, p_N)$ is 
\ben
S(\underline{p}) = \langle {\rm surprise} \rangle = \langle \log \frac{1}{p} \rangle = - \sum_x p_x \log p_x.
\een
$S$ is of course the von Neumann entropy. In a finite quantum system, we can diagonalize the quantum density matrix as in 
\ben
\rho = \sum_x p_x |x\rangle \langle x|
\een
and 
then $S(\rho)$ is given by exactly the same formula or equivalently by
\ben
S(\rho) = -\tr \rho \log \rho.
\een
It frequently happens that our belief about the probability distribution for some events is different from the true probability distribution e.g., we believe a die is fair but it actually isn't. 
Suppose we believe the distribution is $q_x$ but it really is $p_x$. The difference between our average surprise and the actual average surprise is
\ben
S(\underline{p}|\underline{q}) = \langle {\rm our \, surprise} \rangle - \langle {\rm actual \, surprise} \rangle = \langle \log \frac{1}{q} \rangle - \langle \log \frac{1}{p} \rangle =  \sum_x p_x \log \frac{p_x}{q_x}.
\een
$S(\underline{p}|\underline{q})$ is called the {\it relative entropy}, or sometimes the KL divergence, or information gain. It can be seen as the information gained if we believed the distribution was $q_x$ but now we learn it is $p_x$ and is thus a measure of 
distinguishability between the two distributions. It is not symmetric in the two probability distributions as is psychologically plausible if we consider our surprise ($S(\underline{p}|\underline{q}) = \log 2$) at 
learning that the probability distribution for a the events `head' and `tail' of a coin actually is totally unfair $\{0,1\}$ when we believed the coin toss to be fair $\{ \frac{1}{2}, \frac{1}{2}\}$ versus our surprise ($S(\underline{p}|\underline{q})=\infty$) at learning it is fair when we believed it to be totally unfair. Indeed, in the first case, we must see in $n$ drawings the events `tail' $n$ times. Since we believe the coin is fair we ascribe a low but finite probability of $2^{-n}$ to this which never becomes zero, whereas in the opposite case any occurrence of `head' will instantly destroy our belief that the probability distribution was $\{0,1\}$.

More formally, let $\{ p_{\rm head}, p_{\rm tail} \}$ and $\{ q_{\rm head}, q_{\rm tail} \}$ be two probability distributions and
\ben
\begin{split}
C_p =& \bigg\{ (i_1, \dots, i_n) \in \{ {\rm head,tail} \}^n \, : \, \text{relative occurrence of head / tail  is} \\
&\text{$= p_{\rm head}$ / $= p_{\rm tail}$ within small tolerance} \bigg\}.
\end{split}
\een
$C_p$ is a subset of $\{ {\rm head,tail} \}^n$ to which an incorrect belief $\{ q_{\rm head}, q_{\rm tail} \}$ would assign a very small probability whereas it actually has probability tending 
to one for large $n$ and tolerance going to zero. Indeed, one can show
\ben
{\rm Prob}_q(C_p) \sim e^{-nS(\underline{p}|\underline{q})}
\een
for asymptotically large $n$ and small tolerance, which gives another interpretation of $S(\underline{p}|\underline{q})$. 

\medskip
\noindent
{\bf Exercise 2:}
Using that $-\log$ is a convex function, derive Klein's inequality $S(\underline{p}|\underline{q}) \ge 0$. Thus, our surprise is usually larger than the true surprise.

\medskip
\noindent

Contrary to the ordinary von Neumann entropy, there isn't a unique generalization of the relative entropy for 
two non-commuting density matrices $\rho, \sigma$. Some possibilities are:
\begin{itemize}
\item 
Araki-Umegaki: $S_{AU}(\rho | \sigma) = \tr \rho \log \rho - \tr \rho \log \sigma$.
\item 
Measured: For an ONB $\{|x\rangle\}$, the transition amplitudes $p_x = \langle x | \rho | x\rangle, q_x =  \langle x | \sigma | x\rangle$ give 
classical probability distributions. We can define 
\ben
S_M(\rho | \sigma) = \max_{{\rm ONBs} \, \{ |x\rangle \}} S(\underline{p}|\underline{q}), 
\een
which is called the measured relative entropy since we can think of $|i\rangle$ as the eigenvectors of some observable to be measured. 
\item 
Belavkin-Staszewski: $S_{BS}(\rho | \sigma) = -\tr \rho \log (\sqrt{\rho}^{-1} \sigma \sqrt{\rho}^{-1})$. 
\end{itemize}
These are by no means the only possibilities. What separates a good generalization from a bad one is the satisfaction of an important inequality called 
the ``Data Processing Inequality'' (see below) and the existence of an operational meaning. An operational meaning is roughly a quantum information theoretic 
protocol for solving a problem for which the corresponding relative entropy gives the asymptotic rate of success. The Araki-Umegaki entropy 
passes both benchmarks, whereas the measured relative entropy does not satisfy the DPI and no straightforward 
operational meaning for the Belavkin-Staszewski entropy seems to be known. Therefore, in the following, I will restrict attention to the Araki-Umegaki version
which I simply write as $S = S_{AU}$.

Density matrices (mixed states) often arise from pure states by restriction to a subsystem: Let $\sH_{AR} = \sH_A \otimes \sH_R$ be a composite quantum system consisting of $A$
and a reservoir $R$. Let $|\Psi\rangle$ be a pure state of $AR$. Then, 
\ben
\rho_A := {\rm Tr}_R |\Psi\rangle \langle \Psi| 
\een
is called the reduced density matrix. Its von Neumann entropy is often called the {\it entanglement entropy}
\ben
S_{\rm EE}(\Psi) = S(\rho_A).
\een
The entanglement entropy depends on the ``cut'' between the system $A$ and reservoir $B$ and is in this sense subjective.

\medskip
\noindent
{\bf Exercise 3:}
Derive Klein's inequality $S(\rho|\sigma) \ge 0$. Suppose $\sH_{AB} = \sH_A \otimes \sH_B$ is the Hilbert space of a composite system $AB$, with density operator $\rho_{AB}$
and reduced density operator $\rho_A = \tr_B \rho_{AB}$ and similarly for $B$. Show  that 
$-S(\rho_{AB}) + S(\rho_A) + S(\rho_B) = S(\rho_{AB}|\rho_A \otimes \rho_B) \ge 0$. Show that $S(\rho_A)=S(\rho_B)$ if $\rho_{AB}$ is pure hence that 
$S(\rho_{AB}|\rho_A \otimes \rho_B) = 2S(\rho_A)$ in this case.

\section{Quantum channels and DPI}

Before considering quantum channels, we discuss their classical analogue called {\it stochastic matrices}. Consider a system $A$ whose states labelled by $x \in X$, and a system $B$ 
whose states are labelled by $y \in Y$. For simplicity, $X$ and $Y$ are assumed to be finite sets. A {\it stochastic matrix}, $K(y|x)$, by definition has the property that 
whenever $p_x, x \in X$ is a probability distribution, then so is 
\ben
q_y = \sum_x K(y|x) p_x. 
\een
Since we require that $q_y \ge 0$ and $\sum_y q_y = 1$, a stochastic matrix must therefore satisfy $K(y|x) \ge 0$ and $\sum_y K(y|x) = 1$, and vice versa, any matrix with these two properties is stochastic. 

\noindent
{\bf Example 1.}
As our first example, let $X=\{1, \dots, 2N\}, Y=\{1,\dots,N\}$, and let $K(y|x)$ be defined by $q_1 = p_1+p_2, q_2 = p_3+p_4$, and so on. Clearly, $q_y$ is a probability distribution, so 
the $N \times (2N)$ matrix $K(y|x)$ corresponding to this map is stochastic. The map can be considered as a kind of ``coarse graining'' if we imagine that $p_x$ describes the probability of the color ``black'' in an array of $2N$ pixels, and $q_y$ that of the color black in an array of $N$ pixels which have been blocked together pairwise. 

\noindent
{\bf Example 2.}
For our second example we take $X=Y=\{E_1,\dots,E_N\}$, let us associate, for better intuition, each state $x$ with a quantum state $|x\rangle$, and $p_x$ with a density matrix as in
\ben
\rho = \sum_x p_x |x\rangle \langle x|.
\een
Let us imagine that $|x\rangle$ are the eigenstates of some Hamiltonian, $H_0$ and that $H_0$ is offset by a small perturbation by $H_1$ inducing 
transitions between these states per unit time at rate $A_{x,x'} \propto |\langle x| H_1 |x'\rangle|^2$ (Fermi's golden rule). The probabilities $p_x$ will then drift 
according to the differential equation 
\ben
\frac{d}{dt} p_x(t) = \sum_{x':x' \neq x}(A_{x,x'} p_{x'}(t) - A_{x',x} p_x(t)), \quad p_x(0) = p_x.
\een
The map sending $p_x \to p_x(t)$ for a fixed $t \ge 0$ is then a stochastic matrix, $K(t; y|x)$. Actually, by the usual considerations about the uniqueness of initial value problems, this is even a 1-parameter family of stochastic maps 
satisfying the composition law
\ben
K(t+t'; y|x) = \sum_z K(t';y|z) K(t;z|x).
\een

One of the key properties of any stochastic map $K$ is that 
\ben
S(\underline{p}|\underline{q}) \ge S(K\underline{p}|K\underline{q}).
\een
We shall prove this below in the quantum setting, which includes this inequality as a special case. 
Here I only remark that  it is very plausible for the coarse graining example 1, 
since $K$ decreases the distinguishablity between probability distributions. In as far as example 2 is concerned, 
we may observe that $\underline{p}^{\rm eq} = (1,1,\dots,1)/N$ corresponding to the maximally mixed density matrix/micro-canonical ensemble
\ben
\rho^{\rm eq} = \frac{1}{N} \sum_x |x\rangle \langle x|
\een
is an equilibrium distribution, meaning that $K(t) \underline{p}^{\rm eq} = \underline{p}^{\rm eq}$. Hence, in this example, it follows that 
\ben
S(\underline{p}(t_2) | \underline{p}^{\rm eq}) \le S(\underline{p}(t_1) | \underline{p}^{\rm eq}) \quad \text{if $t_2>t_1$.}
\een
This is again plausible, because over time the density matrix is expected to approach the equilibrium state (micro-canonical ensenble).

Consider now a finite quantum system with Hilbert space $\sH_A$. A simple-minded generalization would be to define a quantum channel as a linear operator $\tilde T$
between $B(\sH_A) \to B(\sH_B)$ (the bounded operators) such that if $\rho_A$ is a density matrix, then so is $\rho_B = \tilde T(\rho_A)$. A particular example of 
such a $\tilde T$ is given by any stochastic matrix $K(y|x)$ if we restrict to diagonal density matrices $\rho_A = \sum_x p_x |x\rangle \langle x|$ in a fixed ONB 
$\{|x\rangle\}$ of $\sH_A$, i.e.,
\ben
\tilde T[\rho_A] = \rho_B, \qquad \rho_B = \sum_y q_y |y\rangle \langle y|,
\een
where $p_x$ and $q_y$ are related through $q_y = \sum_x K(y|x)p_x$ and $\{|y\rangle\}$ is a fixed ONB of $\sH_B$.

However, this definition is not really well-motivated, and also misses an important additional condition that we must impose on a channel in the quantum case. 
We therefore proceed differently by considering typical operations in quantum systems. 
The quantum observables are all (say, bounded) operators on $\sH_A$ the collection of which forms an algebra $\cA$. I will frequently call 
the irreducible representation $\sH_A$ the ``fundamental representation''. 
A better way to characterize a channel (at first in the Heisenberg picture) is to say that it's a combination of the following basic operations:
\begin{itemize}
\item Time-evolution: A unitary $u$ from $\cA$ mapping observables $a \to u a u^\dagger$. Here  $a^\dagger$ is the adjoint of an observable, often denoted as $a^\dagger$ in the physics literature.
This gives a linear map $\cA \to \cA$.
\item Projective measurement: A projection $p$ from $\sH_B$ to a subspace $\sH_A$ which takes $a \to pap^\dagger$ and gives a map $\cB \to \cA$, where $\cB$ is the observable algebra on $\sH_B$.
\item Ancillary systems: Let $\sH_R$ be the Hilbert space of a reservoir, with observable algebra $\cR$. Embedding $\sH_A$ into $\sH_B = \sH_{AR} = \sH_A \otimes \sH_R$
takes $a \to a \otimes 1_R$. This gives a map $\cA \to \cA \otimes \cR = \cB$.
\end{itemize}
Here the idea is that we could couple a system $A$ to a reservoir $R$ and effectively design 
a Hamiltonian $h \in \cA \otimes \cR$ of the combined system $B=AR$ in which the time evolution $u = e^{ith}$ takes place for a certain $t$ that we can adjust. 
Afterwards, we measure a state $|\phi\rangle_R$ of system $R$ corresponding to a projection $p = 1_A \otimes |\phi\rangle \langle \phi |_R$ from $\sH_B$ 
onto $\sH_A \otimes |\phi\rangle_R$. 

It can be shown that by combining the above three operations, we get the most general {\it unital completely positive map}. A unital 
completely positive map $T: \cB \to \cA$ is by definition a
linear operator such that if 
\ben
\left(
\begin{array}{ccc}
b_{11} & \dots & b_{1n}\\
\vdots & & \vdots\\
b_{n1} & \dots & b_{nn}
\end{array}
\right) \ge 0 \quad \Longrightarrow
\quad 
\left(
\begin{array}{ccc}
T(b_{11}) & \dots & T(b_{1n})\\
\vdots & & \vdots\\
T(b_{n1}) & \dots & T(b_{nn}) ,
\end{array}
\right) \ge 0
\een
and such that $T(1_B) = 1_A$. Here $b_{ij}$ are from $\cB$, $a_{ij} = T(b_{ij})$ are from $\cA$, and the above matrices are therefore block-matrices with entries in $\cB$ respectively $\cA$. 
The notation $X \ge 0$ for some operator on a Hilbert space means that $\langle \chi | X | \chi \rangle \ge 0$ for 
any state $|\chi\rangle$, so in particular we should have $X^\dagger = X$ (equivalently, all eigenvalues of $X$ are non-negative).
We could have just defined a quantum channel by requiring these conditions, i.e. by requiring it to be a unital complelety positive map. 

\medskip
\noindent
{\bf Exercise 3:}
Let $X \ge Y$ for two self-adjoint operators (meaning $\langle \chi | (X-Y) |\chi \rangle \ge 0$ for all kets $|\chi\rangle$). Show that $Y^{-1} \ge X^{-1}$.

\medskip
\noindent
{\bf Exercise 4:}
Show that for any positive integer $n$, the map $T(b) = n (\tr b) 1_B - b$ is $n$-positive from $\cB \to \cB$, i.e. satisfies the condition for block matrices of up to size $n$, but 
not $n+1$-positive. 

\medskip
\noindent
{\bf Exercise 5:}
For $b \in \cB$ consider the block matrix 
\ben
X= \left(
\begin{array}{cc}
1_B & b \\
b^\dagger & b^\dagger b
\end{array}
\right) .
\een
Sandwiching this block matrix between $(\langle \chi | \, \langle \psi|)$ and $\left(
\begin{matrix}
|\chi\rangle\\
|\psi\rangle
\end{matrix}
\right)$ show that this is $\ge 0$ for any $b \in \cB$.
By applying a 2-positive map $T: \cB \to \cA$ with $T(1_B) = 1_A$ to this matrix, show likewise that $T(b^\dagger b) \ge T(b)^\dagger T(b)$. 

\medskip
Since $T$ acts on observables, we are effectively in the Heisenberg picture but we can 
just as well define the ``dual'' channels $\tilde T$ in the Schr\" odinger picture as acting on density matrices, which is closer to our 
considerations in the commutative case. The relationship between both pictures is as usual
\ben
\tr \rho T(a) = \tr \tilde T(\rho)a,
\een
which defines $\tilde T$ from $T$.

\medskip
\noindent
{\bf Exercise 6:} Show that if $\rho_A$ is a density matrix of system $A$ then $\rho_B = \tilde T(\rho_A)$ is a density matrix for 
system $B$. Show that if $T$ is the time-evolution, projective measurement, and ancillary systems channel, then the dual is given by, respectively:
\begin{itemize}
\item Time-evolution: $\tilde T(\rho) = u^\dagger \rho u$.
\item Projective measurement: $\tilde T(\rho) =  \rho \oplus 0_{A^\perp}$.
\item Ancillary systems: $\tilde T(\rho) = \tr_R \rho$.
\end{itemize}

\medskip

The famous data processing inequality (DPI) states that for any density matrices $\sigma, \rho$ of a system $A$ and any completely positive $T$ from a system $B$ to $A$, 
one has 
\ben
S(\rho | \sigma) \ge S(\tilde T(\rho) | \tilde T(\sigma)). 
\een
We interpret this inequality as saying that the distinguishability of two states can't increase if we pass them through a channel. By applying the inequality twice we can also say that the 
distinguishability remains the same for an invertible channel such as unitary time evolution. Both properties are intuitively very plausible.

\medskip
\noindent
{\bf Exercise 7:} For a density matrix $\rho=\rho_{ABC}$ on a Hilbert space $\sH_{ABC} = \sH_A \otimes \sH_B \otimes \sH_C$
define $\rho_{BC} = \tr_A \rho_{ABC}$ etc. and $\sigma = \rho_A \otimes \rho_{BC}$. Define $\tilde T = \tr_C$. 
Applying the DPI to this situation, derive the strong subadditivity formula
\ben
S(\rho_{ABC}) + S(\rho_C) \ge S(\rho_{AC}) + S(\rho_{BC}).
\een

\medskip

I will now sketch a proof of the DPI, following a strategy invented by Petz, see e.g., \cite{petz} for details on this and on many other 
concepts related to entropy and operator algebras. The proof introduces several ideas that a quite typical for the subject and that are  used when we establishing
improvements of the DPI in connection with the QNEC as described further below.

\medskip
\noindent
{\bf Idea 1: Standard representation.} The first idea is closely related to that of state ``purification''. 
I started from the ``fundamental'' representation of all operators (matrices) $\cA$ on the finite dimensional Hilbert space $\sH_A$. A
general state is a density matrix $\rho$ on $\sH_A$. We can alternatively consider the larger Hilbert space $\sH$ given by $\cA$ itself, with Hilbert-Schmidt inner product
\ben
\langle a_1 | a_2 \rangle = \tr a^\dagger_1 a_2 , \quad a_1, a_2 \in \sH \equiv \cA.
\een
On $\sH$, the observable algebra acts by left multiplication as in 
\ben
L(a)|\zeta\rangle = |a\zeta\rangle. 
\een
We also have right multiplication
\ben
R(a)|\zeta\rangle = |\zeta a\rangle. 
\een
Since right multiplication switches the order of the factors, this is a representation of the so called opposite algebra, which is $\cA$ as a vector space with the product in opposite order. 
Unlike the fundamental representation, the left representation is highly reducible. Another way to say this is that $\cA$, represented on $\sH$ by left multiplication has a large commutant. The commutant $\cA'$ of an algebra 
$\cA$ on a Hilbert space is just the set of all operators commuting with any $a$ from $\cA$ and it is again an algebra. In the present case when $\cA$ acts by left multiplication on $\sH$, the 
commutant $\cA'$ is just the opposite algebra acting by right multiplication. Any density matrix $\rho$ on $\cA$ gives the pure state
\ben
|\sqrt{\rho} \rangle \in \sH \quad \Longrightarrow \quad \langle \sqrt{\rho}\, |a|\sqrt{\rho} \rangle = \tr(\rho a), 
\een
where the scalar product is the Hilbert-Schmidt one. If the density matrix $\rho$ has full rank, then 
\ben
\begin{split}
{\rm span} \{ L(a)|\sqrt{\rho}\rangle \, : \, a \in \cA \} = \sH , & \quad \text{($|\sqrt{\rho}\rangle$ is ``cyclic'')} \\
L(a)|\sqrt{\rho} \rangle = 0 \, \Longrightarrow \, a = 0 ,  & \quad \text{($|\sqrt{\rho}\rangle$ is ``separating'')}
\end{split}
\een
A representation of an operator algebra $\cA$ like the left-representation with a cyclic and separating vector is called a ``standard representation''. 

To illustrate the idea of a cyclic and separating vector I will now argue that if we take $\cA_O$ to be the algebra of all observables in a QFT localized in 
a given finite open diamond region $O$ of Minkowski spacetime, then the vacuum $|\Omega\rangle$ in the full Hilbert space $\sH$ is cyclic and separating. This result is called the 
``Reeh-Schlieder theorem''. Note that by causality 
\ben
(\cA_O)' \supset \cA_{O'}, 
\een
where $O'$ is the causal complement of $O$ (in fact equality holds for 
simply connected regions $O$ such as diamonds -- this is called Haag duality). 

First I show that $|\Omega\rangle$ is cyclic, meaning 
that $a|\Omega\rangle, a \in \cA_O$ spans $\sH$ for our fixed finite diamond $O$. Suppose by contradiction that $|\Omega\rangle$ is not cyclic for some $\cA_O$, so we can find a non-zero vector 
$|\chi\rangle$ such that $\langle \chi | a | \Omega\rangle = 0$ for all $a \in \cA_O$. Consider now $\tilde a$ from $\cA_{\tilde O}$ 
for a diamond $\tilde O$ strictly inside $O$. Therefore, 
\ben
\langle \chi | e^{ixP} \tilde a e^{-ixP} |\Omega\rangle = 0 = \langle \chi | e^{ixP} \tilde a |\Omega\rangle 
\een
for any vector $x$ such that $|x^\mu|$ is sufficiently small for all components (here $P_\mu$ is the energy momentum operator which is the infinitesimal generator of translations in 
the $x^\mu$-direction).  This follows because in such a case, $e^{ixP} \tilde a e^{-ixP}$ is in $\cA_O$ as there is by assumption a finite wiggle room between $\tilde O$ and $O$. In the second 
step I used that $P_\mu |\Omega\rangle = 0$. Consider the right side as a function $f(x)$. It has an analytic continuation to $f(x+iy)$ so long as $y$ is inside the future lightcone because 
the spectrum of $P^\mu$ is in the future lightcone so that $e^{yP} = e^{-y^0 P^0+\dots}$ provides an exponential damping as the negative definite operator $-y^0 P^0$ dominates the $\dots$s. 
Thus, we have a function $f(x)$ which vanishes in some real open neighborhood of 
$x=0$ and which is the boundary value of an analytic function $f(z)$ for ${\rm Im}(z)$ inside the future lightcone. Such a function function must vanish for {\em any} $x$, by a standard result 
in complex analysis of several variables called the ``edge of the wedge theorem''!

Thus, $\langle \chi | e^{ixP} \tilde a e^{-ixP} | \Omega\rangle = 0$ for any $x$, meaning that the $|\chi\rangle$ is not only orthogonal to the span $\tilde a|\Omega\rangle$ with $\tilde a$ from $\cA_{\tilde O}$
but also to the span with $\tilde a$ from any translate of $\cA_{\tilde O}$. All such translates generate by definition all of $\sH$, so $|\chi\rangle = 0$, a contradiction. This shows that 
$|\Omega\rangle$ is cyclic for $\cA_O$. Given that this is the case, let  us show that $|\Omega\rangle$ is also separating for $\cA_O$. So let $a|\Omega\rangle = 0$ for some $a$ from $\cA_O$ and let
$a' \in (\cA_O)'$ be an element in the commutant. Then $aa'|\Omega\rangle = a'a|\Omega\rangle = 0$. By causality, $(\cA_O)' \supset \cA_{O'}$ and since $|\Omega\rangle$ is 
cyclic for $\cA_{O'}$ we know that $a'|\Omega\rangle$ span the entire Hilbert space.
Thus $a=0$.  For a more detailed explanation of this argument aimed at physicists see \cite{witten}.

\medskip
\noindent
{\bf Idea 2: Relative modular operator.}  Given a finite quantum system $A$ with Hilbert space $\sH_A$ and  two density matrices $\rho, \sigma$ with $\rho$ invertible, define
\ben
\Delta_{\sigma,\rho} := L(\sigma)R(\rho^{-1}). 
\een
$\Delta_{\sigma,\rho}$ acts on the standard Hilbert space $\sH$ of the standard representation (defined above, where $L$ and $R$ are the left and right representations) and 
not $\sH_A$ and is called the relative modular operator. 

\medskip
\noindent
{\bf Exercise 8:} Show that $\Delta_{\sigma,\rho}$ is a self-adjoint operator such that $\Delta_{\sigma,\rho} \ge 0$ and such that 
\ben
S(\rho|\sigma) = -\langle \sqrt{\rho}| \log \Delta_{\sigma,\rho}|\sqrt{\rho} \rangle, 
\een
where the scalar product is the Hilbert-Schmidt one. 

\medskip
\noindent
{\bf Idea 3: Petz' operator $V$ and operator monotone functions.}
A very useful trick for deriving inequalities in connection with channels is to introduce a linear operator $V$ defined by
\ben
V b|\sqrt{\rho_B} \rangle := T(b) |\sqrt{\rho_A} \rangle, 
\een
where $\rho_B = \tilde T(\rho_A)$. Note that this $V$ maps vectors in the standard representation of $\cB$ to vectors in the standard representation of $\cA$. 
An important property of $V$ is that it is a contraction meaning it cannot increase the length of a vector that it is acting on relative to the Hilbert-Schmidt inner product:
\ben
\begin{split}
\| V b|\sqrt{\rho_B} \rangle \|^2 =& \| T(b) |\sqrt{\rho_A} \rangle \|^2 \\
=& \tr \sqrt{\rho_A} T(b) ^\dagger T(b) \sqrt{\rho_A} \\
\le & \tr \sqrt{\rho_A} T(b^\dagger b) \sqrt{\rho_A} \\
=& \tr b^\dagger b \tilde T(\rho_A) \\
=& \tr b^\dagger b \rho_B \\
=&\| b|\sqrt{\rho_B} \rangle \|^2
\end{split}
\een
where I used exercise 5 in the key step $\le$.
Contractions have a nice interplay with so-called operator monotone functions, i.e. functions such that $X \ge Y$ for two not 
necessarily commuting non-negative operators implies $f(X) \ge f(Y)$. A function of an operator is defined by first diagonalizing it
as in $X = U^\dagger D U$ and then setting $f(X) = U^\dagger f(D) U$ where $f(D)$ is just the function applied to the diagonal elements of the diagonal matrix $D$.
Applying this condition to two scalars implies that $f$ is monotone in the ordinary sense (non-decreasing), but operator monotonicity is stronger. 
In fact, any operator monotone functions has a representation of the form 
\ben
f(x) = f(0) + xf'(\infty) + \int_{0+}^{\infty-} \frac{x(1+t)}{x+t} w_f(t) dt.
\een
Here $w_f(t) \ge 0$ is an $f$-dependent weight function. Note that a Riemann approximation of the integral $f(X)$ can be seen as a sort of positive linear combination involving the 
the operator valued function $ \frac{X}{X+t} =  \frac{t^{-1}}{t^{-1}+X^{-1}}$ which is operator monotone in $X$ by the next exercise. This is basically a proof why such $f$ are operator monotone. The representation of 
$f$ reduces many statements for operator monotone functions to the corresponding statements for the elementary function $ \frac{1}{1+X^{-1}}$.

\medskip
\noindent
{\bf Exercise 9:} Show by elementary means that $f(x) = (t+x^{-1})^{-1}$ with $t>0$ is operator monotone. Hint: 
For $X \ge Y \ge 0$ consider the interpolating family $C_\lambda = \lambda X +(1-\lambda)Y$ 
and then consider $d/d\lambda f(C_\lambda)$. Show that $f(x) = x^\alpha$ has an integral representation with non-negative weight function $w_\alpha(t)$ if and only if 
$\alpha$ is between $0$ and $1$. Determine this weight function. Likewise, show that $\log x$ is operator monotone. 

\medskip
\noindent
The nice thing about operator monotone functions and contractions is that 
\ben
V^\dagger f(X) V \le f(V^\dagger X V), 
\een
a fact which is relatively easily seen for projections $V=P$ from the above representation of $f$. 

\medskip
\noindent
{\bf Exercise 10:} Show $\| \Delta_{\rho_A, \sigma_A}^{1/2} Vb |\sqrt{\rho_B} \rangle \|^2 \le \| \Delta_{\rho_B, \sigma_B}^{1/2} b |\sqrt{\rho_B} \rangle \|^2$ 
hence that $V^\dagger \Delta_{\rho_A, \sigma_A} V \le \Delta_{\rho_B, \sigma_B}$. 

\noindent
\medskip

Using exercise 10, and the interplay between operator monotone functions and contractions you can see immediately that
\ben
V^\dagger (\log \Delta_{\rho_A, \sigma_A}) V \le \log( V^\dagger \Delta_{\rho_A, \sigma_A} V) \le \log \Delta_{\rho_B, \sigma_B}, 
\een
and then
\ben
\begin{split}
S(\rho_B | \sigma_B) =& -\langle \sqrt{\rho_B}| \log \Delta_{\sigma_B,\rho_B}|\sqrt{\rho_B} \rangle \\
\le& -\langle \sqrt{\rho_B}| V^\dagger (\log \Delta_{\sigma_A,\rho_A}) V|\sqrt{\rho_B} \rangle\\
=&-\langle \sqrt{\rho_A}| \log \Delta_{\sigma_A,\rho_A}|\sqrt{\rho_A} \rangle \\
=& \, S(\rho_A | \sigma_A)
\end{split}
\een
using that $V|\sqrt{\rho_B} \rangle$ is $|\sqrt{\rho_A} \rangle$.
This is the DPI!

\section{Von Neumann algebras}

So far I have been very sloppy about the distinction between observable algebras for finite dimensional quantum systems (i.e. matrix algebras) and 
infinite dimensional operator algebras such as the algebras $\cA_O$ in QFT associated with a diamond $O$. To get a clearer understanding 
we should make some basic assumptions about these algebras. Observables such as quantum fields $\phi(x)$ are not bounded operators
(in fact for a sharp $x$ they are properly speaking not operators at all), but we can average $x$ against some sampling function which is 
nonzero in $O$. This still gives us an unbounded operator, i.e. an operator whose spectral values are unbounded, 
but then we can furthermore take bounded functions of it and if we collect all 
bounded functions of the smeared operators, we get an algebra of bounded operators. 

Motivated by this reasoning I will thus restrict attention to algebras of bounded operators and since we want a notion of self-adjoint operator, 
this algebra should be closed under $a \to a^\dagger$. More specifically, I will from now consider so called von Neumann algebras $\cA$, meaning that:

\begin{itemize}
\item $\cA$ should be a subset of the bounded operators on some Hilbert space $\sH$ which is closed under $\dagger$ and product and which contains 
the identity operator.
\item If $a_n$ is a sequence of elements in $\cA$ and $a$ some bounded operator such that $\langle \chi | (a_n - a) | \phi \rangle  \to 0$ for any pair 
of vectors, then $a$ should also be in $\cA$. 
\end{itemize}

The second requirement in a sense says that if we can reproduce all matrix elements of some operator $a$ to arbitrary precision by 
matrix elements from operators in $\cA$, then $a$ should itself be in $\cA$, which seems a reasonable idealization and is very convenient 
mathematically. A standard text on von Neumann algebras is \cite{takesaki}.

We have seen even for matrix algebras that they can have different representations e.g., the fundamental and the standard 
representation. For general von Neumann algebras, there is no analogue of the fundamental representation except for type I (see below), and thus no 
analogue of a density matrix. The appropriate generalization of a density matrix, i.e. state is that of a positive linear functional, 
meaning a linear functional $\omega: \cA \to \CC$ such that $\omega(a^\dagger a) \ge 0$ for all $a \in \cA$ and $\omega(1) = 1$. Even in the general case there always is an 
analogue of the standard representation, by celebrated work of Haagerup and Araki. If such a standard representation on $\sH$ is chosen, then for every linear functional $\omega$ there is a unique
state $|\sqrt{\omega}\rangle$ in $\sH$ representing it. Of course in the general case, the square root in $|\sqrt{\omega}\rangle$ is purely a notation 
indicating a specific choice of the purification guaranteed by the works of Haagerup and Araki, and its construction is highly non-trivial and beyond the scope of these notes. 

\medskip
\noindent
{\bf Exercise 10:} If $\cA$ is the von Neumann algebra of all bounded operators on the Hilbert space $\sH_A$ (fundamental representation), show that any 
state functional $\omega$ has the form $\omega(a) = \tr \rho a$ for a unique statistical operator $\rho$.  

\medskip
\noindent
Although this is not strictly necessary, I will assume that $\sH$ is a standard representation -- more precisely I assume\footnote{This is not quite the same 
as the technical definition of a standard form of a von Neumann algebra, which always exists but does not require such a cyclic and separating vector.} that there exists 
a separating and cyclic vector $|\Omega\rangle$ in $\sH$. By the Reeh-Schlieder Theorem, we are in a standard representation when $\cA_O$
is the algebra of local observables of some diamond and $|\Omega\rangle$ can be taken, for example, as the vacuum vector. 

It is also convenient
to restrict to von Neumann algebras which are factors, i.e. ones for which $\cA \cap \cA'$ consists only of the identity operator. Since non-factorial
von Neumann algebras can be decomposed as a direct sum (or integral) of factors, this is not really a restriction and the local von Neumann algebras 
in QFT are usually factors anyhow. Von Neumann factors come in three different types (in a standard representation, the type of the commutant $\cA'$
is the same as that of $\cA$):

\begin{itemize}
\item type I: This type is defined by the fact that there exists a (actually many) minimal projection $e \in \cA$, i.e. there is no other non-trivial projection whose 
range is properly contained in that of $e$. Von Neumann factors of type I have a representation as the algebra of all bounded operators 
on some Hilbert space $\sH_A$, the ``fundamental representation'' considered before, but of course we can also consider the standard representation introduced before -- the classification into 
types is independent of the representation. The minimal projections $e = |\phi\rangle \langle \phi|$ project precisely onto the pure states $|\phi\rangle$ 
of the fundamental representation $\sH_A$. Von Neumann algebras of type I are isomorphic if and only if the cardinality of an ONB of $\sH_A$ is the same, so 
we have for instance I$_n$ (when the dimension of $\sH_A$ is $n$).

\item type II: This type is defined by the fact that there no minimal projections but there are so called finite projections $e$. A projection $e$ is 
called finite if there does not exist another projection $f\in \cA$ sucht that $f<e$ and $f = vv^\dagger, e=v^\dagger v$ for some isometry $v \in \cA$. A type II factor 
with a tracial state, i.e. a positive linear functional $\tau$ satisfying $\tau(1) = 1$ and $\tau(a_1a_2)=\tau(a_2a_1)$ for all $a,b \in \cA$, is said to be of type II$_1$.
By a famous theorem of Connes, the type II$_1$ factor is unique up to isomorphism (provided it is ``hyperfinite''\footnote{
A von Neumann algebra $\cA$ is said to be hyperfinite if there exists an increasing sequence $\{ \cN_n \}$ of type I$_n$ algebras 
exhausting $\cA$ (in the sense that $(\cup_n \cN_n)'' = \cA$).
}). Any other type II factor 
is obtained by taking a tensor product of a II$_1$ factor with a type I$_\infty$ factor. These factors are called II$_\infty$.

\item type III: This type is defined by the fact that there is no finite nor minimal projection. There is a finer classification of type III factors into III$_\lambda$, $0 \le \lambda \le 1$ due 
to Connes. This finer classification which uses modular theory (see below) is beyond the scope of these lectures. By a famous theorem of Haagerup, the type III$_1$ factor 
is unique up to isomorphism (provided it is ``hyperfinite''\footnote{
In QFT we expect the algebras $\cA_O$ for a diamond to be hyperfinite due to the split property: Let $O_n$ be an increasing sequence of diamonds 
exhausting $O$ such that the boundaries of $O_n$ and $O_{n+1}$ have a non-zero distance. Then by the split property there exist type I$_\infty$ factors 
$\cA_{O_n} \subset \cN_n \subset \cA_{O_{n+1}}$ in between. Each such type I$_\infty$ factor may moreover be exhausted by type I$_m$ factors.
}
).
\end{itemize}

Also beyond the scope of these lectures is the result that, under fairly natural conditions, the local algebras in QFT $\cA_O$ are hyperfinite factors of type III$_1$ or direct integrals thereof \cite{buchholz}. 
The mentioned uniqueness result  for such factors leads to the philosophically appealing conclusion that the local algebras can be seen as 
avatars of Leibniz' monads, i.e. elementary building blocks of the physical world without internal structure in a sense. The dynamical structure of QFT is determined rather by 
the relationship between the monads, i.e. the relationship between the $\cA_O$'s for different $O$s. For systems involving quantized gravity, algebras of type II have recently been suggested
\cite{Witten1}.

\medskip

To get a first sense for the difference between types II, III and type I, let's see that in types II, III, there is no analogue of the fundamental representation and no pure states. 
Intuitively this is because a pure state corresponds to a rank one projection. A rank one projection $e$ can't have a projection below it different from zero, i.e. it would be a minimal projection, 
which however does not exist for these types. To give a more formal proof I first explain clearly what is a pure state in the general case, since this may seem a representation dependent statement: e.g., for matrix algebras $\cA$, we have seen that a mixed state, i.e. 
density matrix $\rho$ in the fundamental representation may be purified to the pure state $|\sqrt{\rho}\rangle$ in the standard representation. The invariant way to define 
a pure state is to consider a state as a positive linear functional, $\omega$. Then a pure state is one such that we cannot write $\omega = \lambda \omega_1 + (1-\lambda)\omega_2$
as a non-trivial convex linear combination of other states. For matrix algebras (type I$_n$), state functionals correspond to density matrices, and this condition says precisely 
that the density matrix of $\omega$ is rank one. 

Let us now see that a type II or III algebra does not have any pure state functionals $\omega$. As I said, we consider $\cA$ in a standard representation on $\sH$. Let $\omega$ be our putative pure state functional and $|\sqrt{\omega}\rangle$ its canonical vector representative in $\sH$. Consider the closed subspace spanned by all $a|\sqrt{\omega}\rangle$ where $a$ is from $\cA$ and 
let $e'$ be the orthogonal projector onto it. By construction, since the subspace is left invariant by the action of $\cA$, we must have $[a,e']=0$ for any $a \in \cA$, so $e' \in \cA'$. One can see that it is in fact the smallest projection from $\cA'$ such that $\omega'(e') = 1$ (here $\omega'(a') = \langle \sqrt{\omega}| a '| \sqrt{\omega} \rangle$. A type II or III factor has no minimal 
projections so there is a projection $0<f'<e'$ from $\cA'$ and this $f'$ must have $0 < \lambda:=\omega'(f') < 1$. Now we set 
\ben
\omega_1(a) = \langle \sqrt{\omega}| f'af'| \sqrt{\omega} \rangle/\lambda, \quad
\omega_2(a) = \langle \sqrt{\omega}| (e'-f')a(e'-f')| \sqrt{\omega} \rangle/(1-\lambda). 
\een
Then it is easy to see that $\omega_1, \omega_2$ are states on $\cA$, i.e. positive normalized linear functionals each of which is different from $\omega$, and that $\omega = \lambda \omega_1 + (1-\lambda)\omega_2$. So $\omega$ cannot be pure. 

\medskip

Because there isn't a pure state for types II and III, we also cannot have a fundamental (i.e. irreducible) representation. In particular, if $\cA_W$ is the algebra for a wedge and 
$\cA_{W'} = (\cA_W)'$ its commutant, the decomposition of the standard representation $\sH = \sH_W \otimes \sH_{W'}$ such that $\sH_W$ is a fundamental representation of $\cA_W$
can't exist because these algebras are type III, and the reduced density matrix $\rho_W = \tr_{W'} |\Omega \rangle \langle \Omega|$ (or of any other state) can't exist either. This creates a problem 
if we want to compute the von Neumann entropy of this density matrix. 

\medskip

How can we see that a von Neumann factor is type III if we were to meet one? Perhaps the clearest intuition for why local algebras are of type III in QFT is the following 

\medskip
\noindent
{\bf Characterization of type III} \cite{driessler}. $\cA$ is of type III if it is ``infinite'' (the opposite of the confusing terminology ``finite'', meaning in the present context not that it is finite-dimensional but 
that $v^\dagger v = 1 \Longrightarrow vv^\dagger = 1$) and there is a family of ``automorphisms'' $\alpha_n$ (i.e. channels preserving the $*$ and product of $\cA$) and a state $\omega$ such that
\begin{itemize}
\item $\lim_n \langle \psi | \alpha_n(a) |\psi \rangle = \omega(a) \langle \psi | \psi \rangle$,
\item $\lim_n \| [\alpha_n(a)-a,b] \psi \| = 0$,
\end{itemize} 
for any ket $|\psi\rangle$ and $a,b \in \cA$. Here the norm of a vector is $\| \chi \| = \sqrt{\langle \chi | \chi \rangle}$.

\medskip
\noindent
The relation between the above characterization of type III and the original definition is not particularly obvious and goes through modular theory. It is beyond the scope of these notes. 
Let's instead discuss the conditions in the above characterization.
The first condition is a kind of ``ergodic theorem'' because it implies that the average $n^{-1}\sum_{j=1}^n \alpha_j(a) = \omega(a)1$ in the sense of expectation values: ``time average'' $=$ ``ensemble average'', thinking of $\alpha_n$ as implementing time translations. The second condition is a kind of time-like clustering property within this interpretation. 
Of course this interpretation depends on the given meaning of $\alpha_n$ which in turn depends on the context. I personally find it most appealing to think of $\alpha_n$ as scaling the quantum field
arguments by $n^{-1}$ so $\alpha_n$ would be dilatations. Then the first condition says roughly that there are no bounded observables at point $0$ other than multiples of 
the identity and the second condition says that operators whose localization is squeezed to very short distances commutes approximately with any other observable with a 
fixed wavelength on any state with a fixed wavelength. Of course this requires dilatations to be a symmetry of the theory, which is strictly true only in CFTs but morally true for 
any QFT with a UV fixed point or an asymptotically free QFT. At any rate, type III 
has to do with the fact that in a sense there are no local observables associated with a point which in turn is a kind of short distance clustering/decoupling condition. 

\medskip
\noindent
{\bf Example.} As a prototypical example we consider a Majorana fermion on a lightray. The lightray corresponds to the left moving degrees of freedom of a 1+1 dimensional massless 
theory of Majorana fermions. I will show how to construct different types of von Neumann algebras in this example depending on the representation of the basic anti-commutation relations. 
The canonical anti-commutation relations and $\dagger$ are
\ben
\label{CAR}
\{ \psi_{f_1}^\dagger, \psi_{f_2}^{} \} = 2 {\rm Re} \int_{-\infty}^\infty du f_1(u)^* f_2(u) \cdot 1, \quad \psi_f^\dagger = \psi_{f^*}. 
\een
Here $f(u)$ is a sufficiently regular complex valued function of the lightray variable $u$ vanishing rapidly at infinity. 
The relations are a smeared version of the formal relation $\{ \psi(u), \psi(u') \} = \delta(u-u') 1$ and 
$\psi(u)^\dagger = \psi(u)$ under the identification $\psi_f = \int du \psi(u) f(u)$. To obtain a model for a ``wedge algebra'', we 
consider the subalgebra generated by $\psi_f$ where $f(u)=0$ when $u<0$.

To get a von Neumann algebra, we should take an appropriate closure of this algebra. Here are two possibilities:
\begin{itemize}
\item We construct the ``vacuum state'' by saying what is the corresponding state functional $\omega$:
\ben
\omega(\psi_{f_1} \dots \psi_{f_{2n}}) = \sum_\pi \prod_{i=1}^n ( f_{\pi(i)}^*, P_+ f_{\pi(n+i)}^{})
\een 
where $\pi$ runs over all shuffles of $\{1, \dots, 2n\}$ (``Wick's theorem''), and where 
\ben 
P_+f(u) = (2\pi)^{-1} \int_0^\infty dk \, e^{iku} \hat f(k)
\een
is the positive frequency part, which is a projection operator on $L^2$ with inner product 
\ben
(f_1 , f_2) = \int du \, f_1(u)^* f_2(u).
\een
For an odd number of fields we set $\omega=0$. (You may check that $\omega(\psi(u)\psi(u')) = (2\pi)^{-1} (u-u'-i0)^{-1}$ which 
is indeed the correlator of a left moving Majorana field.)
In a standard representation $|\sqrt{\omega}\rangle$ corresponds to the vacuum 
state and the closure of the algebra generated by $\psi_f$ where $f(u)=0$ when $u<0$ is a von Neumann algebra $\cA$ of 
type III, in fact III$_1$. To see this, we apply the above criteria to dilations $\alpha_n(\psi(u)) = \sqrt{n} \psi(n^{-1}u)$. 

\item We construct the ``ceiling state'' by saying what is the corresponding state functional $\tau$:
\ben
\tau(\psi_{f_1} \dots \psi_{f_{2n}}) = \sum_\pi \prod_{i=1}^n \tfrac{1}{2} ( f_{\pi(i)}^*, f_{\pi(n+i)}^{})
\een 
In a standard representation $|\sqrt{\tau}\rangle$ corresponds to a thermal state at infinite temperature 
and the closure of the algebra generated by $\psi_f$ where $f(u)=0$ when $u<0$ is a von Neumann algebra $\cA$ of 
type II$_1$. The ceiling state provides a state with the trace property ($=$ KMS condition at infinite temperature) that such an algebra must have. 
\end{itemize}

\medskip
\noindent
{\bf Exercise 11.} Check that both the vacuum- and ceiling state correlation functions are compatible with the 
canonical anti-commutation relations  \eqref{CAR}. 

\medskip
\noindent
{\bf Exercise 12.} Apply the above criteria to dilations $\alpha_n(\psi(u)) = \sqrt{n} \psi(n^{-1}u)$ in the first case.
Note that $a$ and $b$ are products of $\psi_f$'s, and we may take the $f$'s to be smooth and of compact support 
in $(0, \infty)$ and that we may take $|\psi\rangle$ as $c |\sqrt{\omega} \rangle$ for some local operator $c$
supported in, say, the interval $(1,2)$ (Reeh-Schlieder). The Reeh-Schieder property fails in the second case.

\medskip
\noindent
{\bf Exercise 13.} Verify the trace property for $\tau$. 

\section{Modular operators and relative modular operators}

For type III algebras there is no fundamental (i.e. irreducible) representation, no pure states, and the notion of a density matrix 
does not make sense. Hence, we cannot actually meaningfully define, say, the von Neumann entropy of the non-existent reduced 
density matrix of the vacuum state for a region such as a wedge in Minkowski spacetime or the exterior of a Schwarzschild black hole. 
One may pass to type I algebras by some sort of cutoff, but then the von Neumann entropy diverges as we remove the cutoff. 

While the von Neumann entropy does not make sense for type III algebras $\cA$, the {\em relative} entropy does! This is because we still have 
a notion of relative modular operator. This object $\Delta_{\rho,\sigma}$ was defined above for density matrices which do not exist for type III, 
but there is a way around this. For a general von Neumann algebra in standard form a state is a positive normalized linear functional: $\sigma(a^\dagger a) \ge 0, \sigma(1)=1$. 
We have a canonical purification $|\sqrt{\sigma}\rangle$ in $\sH$ which for type I is indeed just the square root of the density matrix under the identification of functionals
with density matrices as in  $\sigma(a) = \tr a\sigma$. 
Suppose that $\sigma$ is cyclic and separating and $\rho$ an arbitrary state. Define a conjugate linear operator by
\ben
S_{\rho,\sigma} ( a |\sqrt{\sigma}\rangle ) := a^\dagger |\sqrt{\rho}\rangle. 
\een
Here $a \in \cA$ and you can check (exercise) that this definition makes sense because $\sigma$ is cyclic and separating. $S_{\rho,\sigma} $ is not in 
general a bounded operator, which makes it more difficult to handle. But it is not a totally hopeless object because it can be shown to be `closable'
on the domain of vectors $|\psi\rangle = a |\sqrt{\sigma}\rangle$ where $a$ is from $\cA$ (this is a dense subspace of $\sH$). For a closable operator, 
there is always a polar decomposition
\ben
S_{\rho,\sigma} = J \Delta_{\rho,\sigma}^{1/2}. 
\een
Here $J$ is an operator such that $\langle J\psi | J\phi \rangle = \langle \phi | \psi \rangle$ (anti-unitary), $J^2 = 1$. 
$\Delta_{\rho,\sigma} \ge 0$ is a self-adjoint operator. We call it the relative modular operator because for matrix algebras it is the same object as introduced before, as you may check in

\medskip
\noindent
{\bf Exercise 14.} You are invited to work out what $J,\Delta_{\rho,\sigma}$ are concretely for $\cA=$ matrices of size $n \times n$
and $\sH = \cA$ with Hilbert-Schmidt inner product and the left action of $\cA$: $L(a)|\psi\rangle \equiv a|\psi\rangle = |a\psi\rangle$.  Result:
\ben
\Delta_{\rho,\sigma} = L(\rho)R(\sigma^{-1}), \quad J|\psi\rangle = |\psi^\dagger\rangle.
\een
Show that $\Delta^{it}_{\rho,\sigma} a \Delta^{-it}_{\rho,\sigma} = \rho^{it}a\rho^{-it}$ and that $JaJ=a'$ is from $\cA'$ whenever $a$ is from $\cA$.

\medskip
\noindent
The relative modular operator exists even when the density matrices $\sigma,\rho$ do not exist. Intuitively, their non-existence is somehow related to the 
the fact that the would-be density matrices aren't normalizable. But in $\Delta_{\rho,\sigma} = L(\rho)R(\sigma^{-1})$ we have the density matrix and its inverse, 
so somehow the combination exists $\Delta_{\rho,\sigma}$ due to a cancellation of divergences, 
despite the fact that the naive formula $L(\rho)R(\sigma^{-1})$ makes no sense in general. Likewise, the 
relative entropy defined as $S(\rho|\sigma) = -\langle \sqrt{\rho}| \log \Delta_{\sigma,\rho}|\sqrt{\rho} \rangle$ has a chance to exist even though 
each term in the formal expression $S(\rho | \sigma) = \tr \rho \log \rho - \tr \rho \log \sigma$ makes no sense in general. 
Because we proved the DPI using the relative modular operator, it still holds in the general setting of a channel between two general 
von Neumann algebras. 

Consider the vacuum state for a QFT and $\cA_W$ the algebra of observables for a wedge. Let $\omega_W$ be the state functional on $\cA_W$ induced 
from $|\Omega\rangle$, i.e. formally $\omega_W = \tr_{W'} |\Omega \rangle \langle \Omega |$. Taking both states in the relative modular operator to be equal to 
$\omega_W$, we obtain $\Delta_W := \Delta_{\omega_W, \omega_W}$ (when both states are equal we speak simply of the modular operator). 
The Bisognano Wichmann theorem may be restated as saying 
\ben
\Delta_W^{it} = \exp(it K_W), \quad K_W = \text{generator of Lorentz boosts in $1$-direction.}
\een
This formally follows from the exercise 14 but Bisognano and Wichmann gave a rigorous proof \cite{bw}, see \cite{witten} for an exposition aimed at physicists. 
Another aspect of their argument is that $J=$ CPT operator, where `P' means a
reflection of the $1$-direction. In particular, we have $J \cA_W J = \cA_{W'}$ with $W'$ the opposite wedge. Since $J$ always exchanges an algebra with its commutant, 
this gives Haag duality $(\cA_W)' = \cA_{W'}$. Interestingly, Haag duality does not hold for arbitrary regions 
and its failure has very interesting connections with topological effects, order-disorder and 
electro-magnetic duality \cite{cas3}. 

The flow $a \to \Delta^{it}_{\sigma,\sigma} a \Delta^{-it}_{\sigma,\sigma}$ is called the ``modular flow'' for an expectation functional $\sigma$ on $\cA$. As we just noted, this flow corresponds to boosts if $\sigma=\omega_W$ is the restriction of the vacuum state to a wedge algebra $\cA=\cA_W$. If this restriction would make sense as an honest density matrix, then the modular flow would be ``inner'', i.e. implemented by a unitary operator from $\cA_W$ (i.e. $u_W(t) = \omega_W^{it}$). However, by Connes' classification of type III factors, the modular flow can {\em never} be inner, so we see again that this density matrix cannot exist, even though the modular flow exists. 

\section{Area law}

The modular- and relative modular operators are well defined for arbitrary types of von Neumann algebras but they do not help getting a finite von Neumann entropy. The root cause of the problem is, as discussed that the standard Hilbert space $\sH$ does not factorize into $\sH_W \otimes \sH_{W'}$ with $\sH_W$ a hypothetical fundamental (=irreducible) representation of the algebra $\cA_W$ for a wedge. However, something can be done if we place a finite safety corridor between the wedge $W$ and the opposite wedge $W'$:
\begin{figure}[h!]
\centering
\begin{tikzpicture}[scale=.5]
\draw[snake] (-2,5) -- (6,5);
\draw[snake] (6,-3) -- (-2,-3);
\filldraw[color=black, fill=white, thin] (6,-3) -- (10,1) -- (6,5) -- (2,1) -- (6,-3);
\filldraw[color=black, fill=lightgray, thin] (6.5,-2.5) -- (10,1) -- (6.5,4.5) -- (3,1) -- (6.5,-2.5);
\filldraw[color=black, fill=lightgray, thin] (-2,-3) -- (-6,1) -- (-2,5) -- (2,1) -- (-2,-3);
\draw[ thick, dashed] (-2,-3) -- (6,5);
\draw[thick, dashed] (6,-3) -- (-2,5);
\draw (3.7,3) node[anchor=east]{$H^+$};
\draw (3.7,-1) node[anchor=east]{$H^-$};
\draw (3.2,1) node[anchor=east]{\textcolor{red}{$\epsilon$}};
\draw (-1,1) node[anchor=east]{$W'$};
\draw (5,1) node[anchor=west]{$W$};
\draw (8.4,3) node[anchor=west]{$I^+$};
\draw (8.4,-1) node[anchor=west]{$I^-$};

\end{tikzpicture}
\end{figure}
If the size of the safety belt is $\textcolor{red}{\epsilon} > 0$ between $W$ and $W'$, then  it turns out the 
the tensor product of the reduced vacuum states $\omega_W \otimes \omega_{W'}$ is well defined on the algebra
$\cA_{WW'}$ of the wedge and the shifted opposite wedge, and 
one can see that 
\ben
\label{area}
S(\omega_{WW'}|\omega_{W'} \otimes \omega_W) \sim \frac{{\rm horizon \, area}}{\epsilon^2},
\een
similar to the Bekenstein-Hawking formula but with a renormalized Newton constant $G_{\rm eff} \sim \epsilon^2$. 
In the limit as $\epsilon \to 0$, the left side formally goes over to $2 S(\omega_W)$, giving a quantitative 
version of the divergence of the von Neumann entropy. 

The mathematical reason why the belt helps is the so called split property, which says that, for a finite $\epsilon > 0$, there exists a type I factor 
in between $\cA_W$ and $(\cA_{W'})'$. The form of the leading divergence in \eqref{area} is sometimes referred to as the ``area law''. More precisely, 
we think of  $S(\omega_{WW'} |\omega_{W'} \otimes \omega_W)$ as a regulated version of $2 S(\omega_W)$ (by exercise 3) with length cutoff $\epsilon^2$. 

To get an idea where the area law comes from it is instructive to define the so-called ``distillable entanglement'' between two systems $A,B$. A 
state $\omega_{AB}$ is said to be distillable if there exists a sequence of separable operations $\{ F^{(N)} \}$ of $N$ copies of the system 
$\cA_{AB} = \cA_A \otimes \cA_B$ -- ``separable'' meaning that $F^{(N)} = \sum T_A^{(N)} \otimes T_B^{(N)}$ for some channels $T_A^{(N)} : \cA_A^{\otimes N} \to M_{n_N}({\mathbb C})_A$
and similar for $B$ -- such that (with $\tilde F^{(N)}$ the dual ``Schr\" odinger picture'' channel)
\ben
N \to \infty: \qquad \| \, \tilde F^{(N)} \omega_{AB}^{\otimes N} - |\beta_{n_N}\rangle \langle \beta_{n_N}| \,  \| \to 0, 
\een
where $|\beta_n\rangle = \frac{1}{\sqrt{n}} \sum_{i=1}^n |i\rangle_A |i\rangle_B$ is the maximally entangled state for $M_{n}({\mathbb C})_A \otimes M_{n}({\mathbb C})_B$
and where the norm is the ``trace distance'' between two states. Thus $F^{(N)}$ extracts about $n_N$ entangled pairs from $N$ copies of $\omega_{AB}$. The asymptotic rate of 
distillation is the ``distillable entanglement'':
\ben
D(A,B) =  \lim_{N \to \infty} \sup_{\{ F^{(N)} \}}   \frac{\log n_N}{N}
\een
where we optimize over all distillation strategies. 

Consider now $A$ and $B$ as two disjoint regions of a time zero slice separated by a safety corridor of width $\epsilon$, 
and place subsystems $A_i B_i$ near the corridor as in the following diagram. 
\begin{figure}[h!]
\begin{center}
\begin{tikzpicture}[square/.style={regular polygon,regular polygon sides=4}]
            \draw[thick,blue] ([shift=(60:1cm)] 0, 0) arc (60:120:3cm);
            \draw[thick,red] ([shift=(60:1.2cm)] 0, 0) arc (60:120:3.2cm);

        \filldraw[fill=blue!20!white, draw=blue!50]
        ([shift=(60:1cm)] 0, 0) arc (60:120:3cm) -- (-1.5,0) -- (-0.5, 0) -- cycle;
        \filldraw[fill=red!20!white, draw=red!50]
        ([shift=(60:1.2cm)] 0, 0) arc (60:120:3.2cm) -- (-2.0,2.5) -- (0.0, 2.5) -- cycle;

        \node at (-1.0, 1.0) [square,draw, rotate=0,fill=blue] (a0) {};
        \node at (-1.5, 0.95) [square,draw, rotate=11,fill=blue] (a1) {};
        \node at (-0.5, 0.95) [square,draw, rotate=169,fill=blue] (a2) {};

        \node at (-1.0, 1.8) [square,draw, rotate=0,fill=red] (b0) {};
        \node at (-1.65, 1.72) [square,draw, rotate=11,fill=red] (b1) {};
        \node at (-0.35, 1.72) [square,draw, rotate=169,fill=red] (b2) {};

        \draw (a0) -- (b0);
        \draw (a1) -- (b1);
        \draw (a2) -- (b2);

        \node at (-2, 1.21) {$\varepsilon$};
        \node at (-1.0, 2.3) {\color{red}{$B_{i}$}};
        \node at (-2.5, 2.25) {\color{red!50}{$B$}};
        \node at (-1.0, 0.5) {\color{blue}{$A_{i}$}};
        \node at (-2.5, 0.4) {\color{blue!50}{$A$}};
\end{tikzpicture}

\end{center}
\caption{The the sets $A_i,B_i$ in $d > 2$ spacetime dimensions.}
\label{fig:aibi1}
\end{figure}
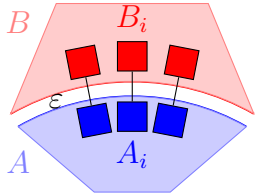
Then we can say that 
\ben
S(\omega_{AB} |\omega_A \otimes \omega_B) \ge D(A, B) \ge D(\cup A_i, \cup B_i) \ge \sum D(A_i, B_i) = n D(A_1, B_1), 
\een
where $n$ is the number of non-intersecting dumbbells (or ``bitstrings'') that we can place across the corridor. In the last 
step we used that by translation and rotation invariance of the vacuum state $\omega$, all dumbbells have an equal distillable entropy, in the second-to-last step 
we used a known super-additivity property of $D$ and in the first step we used the  fact that the relative entropy dominates $D$. Neither property is
particularly obvious; see \cite{Sanders} for the details of the above argument. Consider now reducing the size $\epsilon$ of the corridor to zero while at the same time 
self-similarly shrinking the dumbbells. In a conformally invariant or asymptotically free theory $D$ of a single dumbbell should not 
change or approach a limit when we shrink it, while it is geometrically clear that the number of pairs that we can place across the corridor scales as 
$n \sim \frac{{\rm area}(\partial A)}{\epsilon^{2}}$, so 
\ben
S(\omega_{AB} |\omega_A \otimes \omega_B) \ge D_1 \frac{{\rm area}(\partial A)}{\epsilon^{2}}
\een
in $4$ dimensions, where the area of the boundary of $A$  is = that of $B$ in the limit. 
The constant of proportionality is $D_1:=\lim_{\epsilon \to 0} D(A_1, B_1)$ which we expect in an asymptotically free theory like $SU(N_c)$ pure Yang-Mills in $d=4$, 
to be a universal number times the number $N_c^2$ of asymptotically free gauge fields in the theory. One can show \cite{Sanders} that the universal number which is equal to 
that in free field theory, is non-zero using the Reeh-Schlieder theorem. The number should also not really depend on taking $\omega$ to be the vacuum.

Thus we get the area law for essentially any state in the QFT -- actually just a lower bound but we can also get an upper bound by a different method 
of estimation.  

\section{ANEC}

In classical GR, the null energy condition (NEC) is the statement that the stress energy tensor satisfies
\ben
\Theta_{kk} \equiv \Theta_{\mu\nu} k^\mu k^\nu \ge 0
\een
for any null vector $k^\mu$. The Raychaudhuri equation (the $kk$-component of the Einstein equation) shows that 
a positive $\Theta_{kk}$ tends to make null geodesics focus more and more, i.e. it makes gravity attractive.

In QFT, operators at a sharp point $x$ have an infinite variance and this variance can be used to produce for any given $x$ states 
with $\langle \Psi | \Theta_{kk}(x) | \Psi \rangle < 0$ even in theories where the NEC holds classically. 
In fact, not only that: We can make this expectation value as negative as we like by a suitable choice of $|\Psi\rangle$. 
The following exercise shows that this can be seen as an interference effect.

\medskip
\noindent
{\bf Exercise 15.} (see \cite{helffer}) Consider the following superposition of the vacuum and a 2-particle wave packet in the free KG QFT of mass $m$:
\ben
|\Psi\rangle = \psi_0 |\Omega\rangle + \int d^3p_1 d^3 p_2 \, \psi_2(\vec{p}_1,\vec{p}_2) |\vec{p}_1,\vec{p}_2\rangle.
\een
$\psi_0, \psi_2$ are  such that $|\Psi\rangle$ is normalized. In this theory, $\Theta_{\mu\nu}$, the renormalized (i.e. normal ordered in flat spacetime) stress tensor, is
\ben
\Theta_{\mu\nu} = : \partial_\mu \phi \partial_\nu \phi - \frac{1}{2}\eta_{\mu\nu}(\partial^\sigma \phi \partial_\sigma \phi + m^2 \phi^2):
\een
$\omega_p = \sqrt{\vec{p}^2+m^2}$ is the energy of a particle in this theory, and $|\vec{p}_1,\vec{p}_2\rangle$ is a 2-particle momentum eigenstate.
The exercise is to show that $\psi_0, \psi_2$ can be tuned such that $\langle \Psi | \Theta_{00}(x) | \Psi \rangle < 0$ for, say, $x=0$. You must first rewrite $\Theta_{00}$ 
in terms of creation and annihlation operators $[a(\vec{p}), a(\vec{k})^\dagger] = (2\pi)^3 2\omega_p \delta^3(\vec{p}-\vec{k})$. For this, use the representation
\ben
\phi(x) = \int \frac{d^3 k}{(2\pi)^3 2\omega_k} a(\vec{k}) e^{-i\omega_k t+i\vec{k}\vec{x}} + {\rm h.c.} 
\een
and that $|\vec{p}_1,\vec{p}_2\rangle = a^\dagger(\vec{p}_1) a^\dagger(\vec{p}_2)|\Omega\rangle$. Following \cite{helffer}, one may adopt the following ansatz for $\psi_2(\vec{p}_1,\vec{p}_2)$ you should make an ansatz 
\ben
\psi_2(\vec{p}_1,\vec{p}_2) = \chi(\vec{p}_1+\vec{p}_2) (p_1 p_2)^{\nu-(1/2)},
\een
where $\chi(\vec{p})=\chi_0$ for $p=|\vec{p}|<p_0$ and  $\chi(\vec{p})=0$ otherwise. You should also take $\psi_2(\vec{p}_1,\vec{p}_2) =0$ unless 
$p_1,p_2 < \Lambda$.
This ansatz containing the 4 parameters $\nu,\chi_0,\Lambda,p_0$ represents a superposition of two wave packets with approximately opposite momenta. 
Show that $\langle \Psi | \Theta_{00}(x)| \Psi \rangle$ as a function of $x$ has a pattern of spacelike interference fringes where positive values alternate with negative ones.
By tuning $\nu,\chi_0,\Lambda,p_0$ appropriately, one may show that $\langle \Psi | \Theta_{00}(0) | \Psi \rangle$ may be made arbitrarily negative \cite{helffer}. 

\medskip
\noindent
Exercise 15 suggests that negative values of the expected stress energy tensor perhaps generally alternate with positive values. It is therefore not unreasonable to conjecture that the 
averaged NEC (ANEC) might hold
\ben
\int \langle \Psi|\Theta_{kk}|\Psi\rangle d \lambda \ge 0 \qquad \text{(any $|\Psi\rangle$)}
\een
where the integral is along any null ray with tangent $k^\mu$ and affine parameter $\lambda$. The ANEC is expected to play an 
important role in quantum singularity theorems based on the semi-classical Einstein equation
\ben
\label{sclass}
G_{\mu\nu} = 8\pi G \langle \Psi | \Theta_{\mu\nu} | \Psi \rangle
\een
see \cite{anec4}. Other kinds of averaged energy conditions have also been investigated, see e.g., \cite{anec1,anec2,anec3,anec5}.

How might one prove the ANEC, at least in Minkowski spacetime? 
An important observation by Borchers, Wiesbrock and others \cite{hsm1,hsm2,hsm3,hsm4} from the 1990s is that we can get  interesting operators from {\em two} nested wedges:
\begin{center}
\begin{tikzpicture}[scale=1, transform shape]
\filldraw[color=gray] (0,0) -- (3,3) -- (6,0) -- (3,-3);
\filldraw[color=lightgray] (1,1) -- (3,3) -- (6,0) -- (4,-2);
\draw[black,->] (0,0) -- (3.3,3.3);
\draw[black,->] (0,0) -- (3.3,-3.3);
\draw (4,1.5) node[left]{$W_T$};
\draw (0.5,.9) node[left]{\textcolor{black}{$T$}};
\draw (3.1,3.2) node[left]{\textcolor{black}{$x^+$}};
\draw (3.1,-3.2) node[left]{\textcolor{black}{$x^-$}};
\end{tikzpicture}
\end{center}

The second wedge is 
\ben
W_T = \{x^+ > T(z^A) > 0, x^->0\}, 
\een
where $z^A=(z^1,z^2)$ are coordinates along the edge of the wedge, see the next figure for a 3D illustration.

\begin{figure}[h!]
\begin{center}
  \includegraphics[width=0.6\textwidth,]{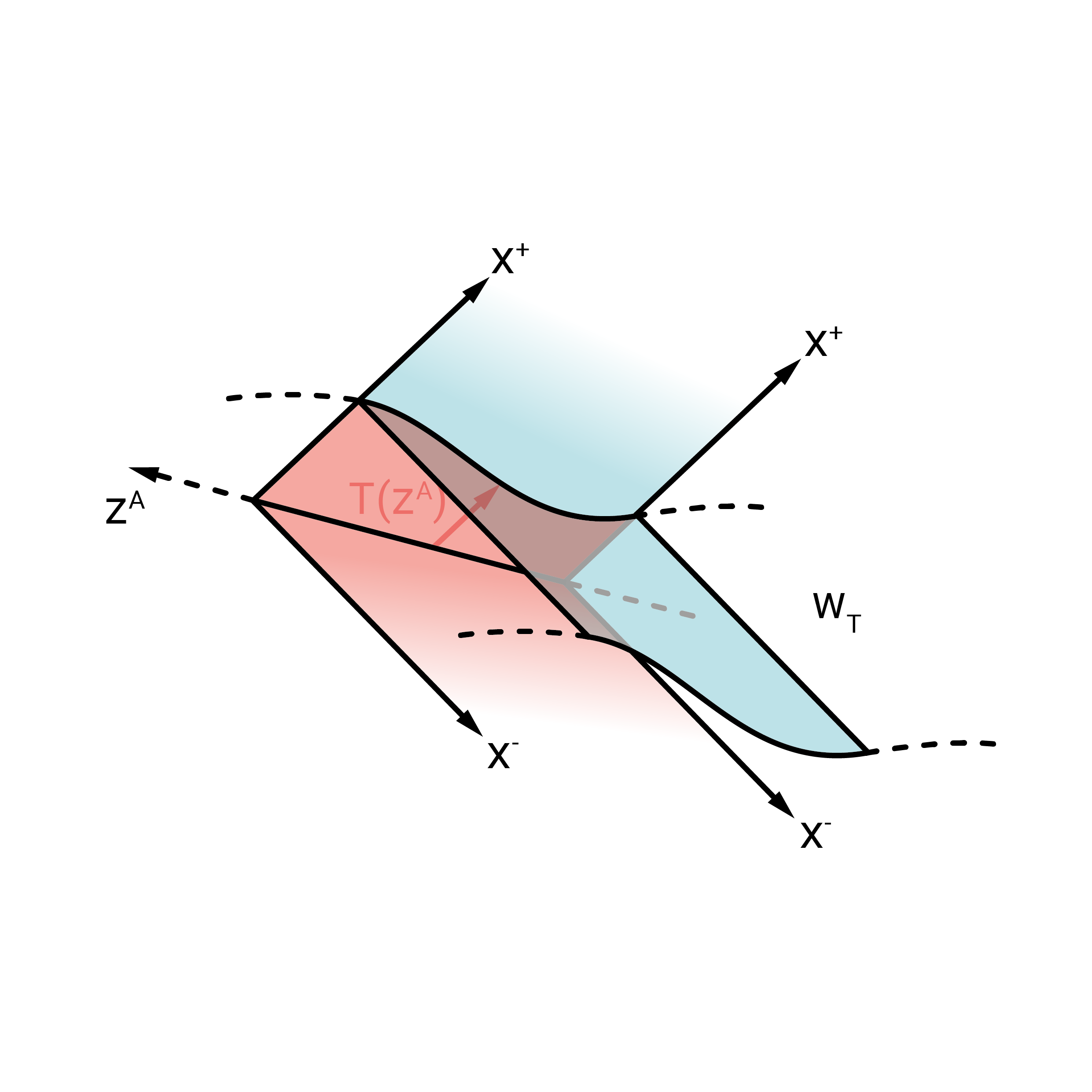}
  \end{center}
  \label{fig:rb}
  \caption{The shifted wedge $W_T$ is in blue, the original wedge in red. The edge of the shifted wedge is parameterized by a function $T(z^A)$.}
\end{figure}

So we have {\em two} modular operators: One for $W$ and one for $W_T$. We know from Bisognano-Wichmann 
that $\Delta_W^{it}$ generates boosts, so points in $W_T$ get moved to points inside $W_T$ for positive values $t>0$ of the parameter. 
This situation is referred to as a ``half sided modular inclusion'' (modulo some other technical requirements \cite{hsm}). If we have a half sided modular inclusion, then we 
get an interesting interplay between $\Delta_W$ and $\Delta_{W_T}^{it}$; in fact one can show:
\begin{itemize} 
\item $\Delta_{W_T}^{it} \Delta^{-it}_W = e^{i(e^{2\pi t}-1) \cE_T}$, 
\item $\cE_T \ge 0$ as an operator,
\item The $\cE_T$'s and boosts generate the Bondi-Metzner-Sachs group,
\item $\cE_T$ is the angle-averaged ANEC operator 
\ben
\label{ETdef}
\cE_T =   \int \left( \int_{-\infty}^\infty   \Theta_{++}(x^+, x^-=0,z^A)  dx^+ \right) T(z^A) \, d^2z, 
\een
\item Since $\cE_T \ge 0$ for all $T>0$, the ANEC holds!
\ben
{\rm ANEC}(z^A) =  \int_{-\infty}^\infty   \Theta_{++}(x^+, x^-=0,z) dx^+ \ge 0
\een
\end{itemize}
The first relation may look mysterious but it is just the relation for two boosts in a pair of coordinate systems related by a null translation. The BMS algebra 
reflects this and the fact that $T(z^A)$, the amount of the null translation, may depend on the coordinates $z^A$ of the edge of the wedge.

Another way to formally state the above relations is to say that the vacuum reduced  to $W_T$ has the density matrix
\ben
\label{remat}
{\rm Tr}_{W_T'} |\Omega\rangle \langle \Omega| \,\, \propto \, e^{-2\pi K_T},
\een
where 
\ben
\label{locboost}
K_T = \int\limits_{x^+ > T(z^A)} [x^+ - T(z^A)] \Theta_{++}(x^+,x^-=0,z^A) dx^+ d^2z + \dots, 
\een
where the terms indicated by dots stand for a contribution involving an integral of a null component of the stress tensor over $I^+$. They must in general be included 
because the surface $\{(x^+,0,z^A) \ : \ x^+ > T(z^A)\}$ by itself is in general not a Cauchy surface of $W_T$, so we must include $I^+$, too.

The connection between $\cE_T$ and the ANEC was in fact recognized only many years after the work by Borchers, Wiesbrock and others, see \cite{casini1}. 
The ANEC is a very strong non-perturbative constraint on matrix elements of the stress energy operator. 
An alternative proof using traditional high-energy methods can also be given in CFTs \cite{hartman}. 
The fact that matrix elements of the ANEC operator have signs give constraining 
relations in conformal bootstrap approach to CFTs; for such topics see e.g., \cite{bs1,bs2}.

\section{QNEC}

A refinement of the ANEC is the so called quantum null energy condition (QNEC), which itself is a limiting case of the even more general ``quantum focussing conjecture'' \cite{qnec1} (not discussed here). The statement is that 
\ben
2\pi \langle \Theta_{kk} \rangle \ge S''_{\rm EE}
\een
where $S''_{\rm EE}$ is the second ``shape variation'' of the entanglement entropy defined with respect to an entangling cut on a null-sheet 
with $k^\mu$ tangent to the affinely parameterized null-generators of the sheet. $S''_{\rm EE}$ nor its integral along a null generator has a sign in general, so 
the QNEC is strictly speaking not stronger than the ANEC.

We now give a more precise discussion of the QNEC in the context where the null sheet is the future horizon of a wedge $W$ in Minkowski spacetime. In this case 
$k^\mu \propto \delta^\mu_+$ in the double null coordinates $(x^+,x^-,z^A)$, where $z^A = (x^2, x^3)$ are coordinates parameterizing the edge of the wedge. 
From a formal viewpoint, the QNEC is discussed most conveniently in terms of relative entropy between a state $|\Psi \rangle$ and the vacuum $|\Omega \rangle$ for the shifted wedge $W_T$ defined as in the previous section, see figure \ref{fig:rb}. Let us 
denote this relative entropy by
\ben
S_{\rm rel}[T] := S(\Psi | \Omega)_{W_T}. 
\een
Since the relative entropy is monotonic (applying the DPI to the inclusion channel from $W_{T_1} \to W_{T_2}$), it follows that 
\ben
S_{\rm rel}[T_2] \le 
S_{\rm rel}[T_1]  \quad \text{if $T_2(z^A) \ge T_1(z^A)$.}
\een
We may also write this infinitesimally as 
\ben
\frac{d}{da}S_{\rm rel}[aT]_{a=0} \le 0.
\een
The QNEC can be seen as the statement about the {\it second} derivative:
\ben\label{qnec}
\frac{d^2}{da^2}S_{\rm rel}[aT]_{a=0} \ge 0.
\een
Since this is supposed to hold for {\it any} regular function $T(z^A) \ge 0$ of the edge coordinates $z^A$, this is in fact not one but many inequalities, which could also be expressed by taking functional derivatives with respect to $T(z^A)$ (``shape variations'') as in the original references, but we will not do this yet. 

For better intuition, it is more instructive to write the QNEC in terms of the ordinary entanglement entropy rather than relative entropy, i.e. in terms of
\ben
\label{relent}
S_{\rm EE}[T] = -{\rm Tr}(\psi_T \log \psi_T).
\een
For this, we first formally introduce the reduced density matrices 
\ben
\omega_T = {\rm Tr}_{W_T'} |\Omega\rangle \langle \Omega| \ , \quad
\psi_T = {\rm Tr}_{W_T'} |\Psi\rangle \langle \Psi| , 
\een
so that formally, 
\ben
\label{relent}
S_{\rm rel}[T] = {\rm Tr}(\psi_T \log \psi_T - \psi_T \log \omega_T).
\een
By substituting into this expression for the relative entropy the relations \eqref{relent}, \eqref{remat} and \eqref{locboost}, we find first that
\ben
\label{relent1}
S_{\rm rel}[aT] = -S_{\rm EE}[aT]  + \,\,\, 2\pi \!\!\!\!\!\! \int\limits_{x^+ > aT(z^A)} [x^+ - aT(z^A)] \langle \Theta_{++}(x^+,x^-=0,z^A)\rangle_\Psi \, dx^+ d^2z + \dots
\een
The dots, which correspond to the terms in \eqref{remat} not written out disappear when we take a derivative with respect to $a$, indicated by a prime $\prime$:
\ben
\label{relent2}
S_{\rm rel}'[aT] = -S_{\rm EE}'[aT]  - \,\,\, 2\pi  \!\!\!\!\!\! \int\limits_{x^+ > aT(z^A)} T(z^A) \langle \Theta_{++}(x^+,x^-=0,z^A)\rangle_\Psi \, dx^+ d^2z .
\een
Taking one more derivative with respect to $a$ gives:
\ben
\label{relent3}
S_{\rm rel}''[aT] = -S_{\rm EE}''[aT]  + \,\,\, 2\pi  \int T(z^A) \langle \Theta_{++}(x^+=aT(z^A),x^-=0,z^A)\rangle_\Psi \, d^2z .
\een
Using at this stage that $S_{\rm rel}''[T] \ge 0$ gives the (formally) equivalent form of the QNEC:
\ben
\label{relent4}
\frac{d^2}{da^2} S_{\rm EE}[aT]_{a=0}  \le \,\,\, 2\pi  \int T(z^A) \langle \Theta_{++}(0,0,z^A)\rangle_\Psi \, d^2z .
\een
Since this holds for all $T(z^A)$, we may be tempted to formally put $T(z^A) = \delta^2(z^A-z^A_0)$, yielding a point-wise condition on the 
expected stress energy tensor, of the sort
\ben
\label{relent4}
\frac{d^2}{da^2} S_{\rm EE}[a\delta_{z_0}]_{a=0}  \le 2\pi  \, \langle \Theta_{++}(0,0,z_0^A)\rangle_\Psi .
\een
The left side is a formal definition of the second shape variation already discussed. 


We now briefly sketch the idea for the proof of the QNEC by \cite{faulkner1,HL25b} (for a heuristic argument, see \cite{qnec2}). 
Both proofs are fundamentally based on a variational principle for the {\it first} 
derivative of the entanglement entropy due to \cite{W1}, and reformulated in terms of the relative entropy as \cite{faulkner1}
\ben
2\pi  \inf_{u' \in W'_{aT} \ {\rm unitary}}  \langle u'\Psi |  \mathcal{E}_{T} |u'\Psi \rangle = -S_{\rm rel}'[aT]
\een
where  ${\mathcal E}_{T}$ is the smeared ANEC operator \eqref{ETdef} appearing in the context of half-sided modular inclusions.
This variational principle immediately shows that $S_{\rm rel}'[bT] \ge S_{\rm rel}'[aT]$ when $b \ge a (\ge 0)$ because 
in the formula for $-S_{\rm rel}'[bT]$ we minimize over a smaller set of unitaries $u' \in \A_{bT}'$ than in the formula for 
$-S_{\rm rel}'[aT]$, where $u' \in \cA_{bT}' \supset \cA_{aT}'$. This demonstrates 
the QNEC in the form \eqref{qnec} but of course we must still establish the variational principle, which is the hard part and which we will not 
try here. The variational principle can also be used to formulate a strengthened form of the QNEC which implies the 
ANEC.

For a general null geodesic in an arbitrary curved spacetime or a congruence representing an outgoing light-sheet as in the QNEC, it is not clear to what extent
 the ANEC nor the QNEC hold on a fixed curved spacetime. But one 
expects that they are restored, modulo terms involving the expansion and shear, 
once backreaction effects are taken into account. It seems plausible that the ANEC and QNEC are a fundamental property of semi-classical gravity. 

\section{Petz map, state recovery and improved DPI}

The QNEC suggests that we should take a closer look at 
\ben
S_{\rm rel}[T=0]-S_{\rm rel}[T]
\een
for a finite shift $T(z^A)>0$. If this quantity is small then we should learn something about the flux of the 
ANEC operator; basically this flux should be small and we should be able to recover the reduced 
density matrix of $|\psi\rangle$ to $W$ from the reduced density matrix of the shifted wedge $W_T$. 
This hints at a connection to the important topic of ``state recovery''. State recovery asks the following question:
Suppose we have a channel $\alpha: \cA_B \to \cA_A$ with dual channel $\tilde \alpha$ giving a state $\rho_B = \tilde \alpha(\rho_A)$ for system $B$ for 
each input state $\rho_A$ from $A$. In which sense can this be ``approximately inverted''? 
In other words, we ask for a channel $R_\alpha: \cA \to \cB$ such that $\rho_A$ and $\tilde R_\alpha( \rho_B )$
are ``close'' for a class of states? Typically, we think of $\tilde \alpha$ 
as a noisy channel erasing some information, and we think of $\tilde R$ as the recovery channel. 
Not always can we recover the initial $\rho_A$ from $\rho_B = \tilde \alpha(\rho_A)$, but we should be able to do so approximately if 
only a small amount of information is erased relative to some reference state. 

In the classical context we have a commutative algebra $\cA_A$ of functions of a random variable $x$ (multiplication operators), and a density matrix corresponds to a probability distribution $p_A(x)$
via $\rho_A = \sum_x p_A(x) |x\rangle \langle x|$. As we have already said, under these identifications, 
a (dual) channel $\tilde \alpha$ from $A$ to $B$ acts on probability distributions by
\ben
p_B(y) = \sum_x K_\alpha(y|x) p_A(x) 
\een
where the kernel $K_\alpha(y|x)$ is a stochastic matrix. So the output state is 
$\rho_B = \sum_y p_B(y) |y\rangle \langle y|$.

A natural guess for defining the recovery channel, $K_R(x|y)$, which is inspired by Bayes' rule is
\ben\label{Bayes}
K_R(x|y) = \frac{K_\alpha(y|x) p_A(x)}{p_B(y)}
\een
wherein $p_A$ is supposed to be a given reference probability distribution which in the quantum case will be a reference state for system $A$. 
We interpret $K_\alpha(y|x)$ as the conditional probability of $y$ given $x$ and then the formula states Bayes' theorem giving the conditional 
probability of $x$ given $y$.

\medskip
\noindent
{\bf Exercise 16.} Suppose we have a channel $\alpha: \cA_B \to \cA_A$ with dual channel $\tilde \alpha$ and a reference state $\sigma_A$ on $A$. 
Then $\sigma_B = \tilde \alpha(\sigma_A)$ is a reference state on $B$. We assume that $\cA_A$ and $\cA_B$ are finite-dimensional matrix algebras
such that the states are represented by density matrices. Check that the ``Petz'' recovery channel (for more on this see \cite{petz}):
\ben
\tilde P_\alpha( \, . \, ) := \sqrt{\sigma_A}\alpha[ \sqrt{\sigma_B}^{-1} ( \, . \,) \sqrt{\sigma_B}^{-1} ] \sqrt{\sigma_A}
\een
acting on density matrices for $B$ reduces to Bayes' formula \eqref{Bayes} in the classical case (with $\sigma_A = \sum_x p_A(x) |x\rangle \langle x|$ etc.). Show that $\tilde P_\alpha$ exactly recovers
$\sigma_A$ from $\sigma_B$ and that, in general $\tilde P_\alpha(\rho_B)$ is a density matrix for system $A$ (which may or may not exactly recover $\rho_A$). 

\medskip
\noindent
The Petz recovery channel described in exercise 16 is of course not the only non-commutative generalization of Bayes' formula \eqref{Bayes}. 
What we want on general grounds is a recovery channel having the property that if, for some reference state $\sigma$ (the vacuum in our case), 
the information loss $S(\rho_A | \sigma_A) - S(\tilde \alpha(\rho_A)|\tilde \alpha(\sigma_A))$ is small, then the difference between $\rho_A$ and $\tilde R_\alpha \tilde \alpha( \rho_A)$ is 
small in some sense. Given that the algebras of observables that we wish to deal with are of type III, we should also have definitions that work for general type. 

The first task is relatively easy. Given $\sigma_A$ and $\sigma_B := \tilde \alpha(\sigma_A)$ we first define the ``KMS inner product'' on $\cA_A$: 
\ben
\langle a_1, a_2 \rangle_\sigma= \langle \sqrt{\sigma_A} | a_1^\dagger \Delta_{A}^{1/2} a_2 |\sqrt{\sigma_A} \rangle \quad 
(= \tr(a_1^\dagger \sqrt{\sigma_A} a_2 \sqrt{\sigma_A}) \quad \text{when $\cA_A =$ type I})
\een
Here $\Delta_A = \Delta_{\sigma_A,\sigma_A}$ is the modular operator for $\sigma_A$.
We do the same for $B$ using $\sigma_B$. So we have KMS inner products on $\cA_A$ and $\cA_B$ and then via  inner products, we 
can define an adjoint (depending on $\sigma_A$) $\alpha^+: \cA_A \to \cA_B$ of the channel $\alpha: \cA_B \to \cA_A$. 

\medskip
\noindent
{\bf Exercise 17.} Check that the Petz recovery channel $P_\alpha$ in exercise 16 is precisely $\alpha^+$.

\medskip
\noindent
We can also define the ``rotated Petz map''
\ben 
P_{\alpha,t}= Ad(\Delta^{-it}_{B}) \circ \alpha^+ \circ Ad(\Delta^{it}_{A})
\een 
where ``Ad'' means the Heisenberg time evolution $Ad(X)Y=XYX^{-1}$. Even more generally, we 
can define the ``twirled Petz map'' by averaging this against some probability density $w(t)$,
\ben 
R_{\alpha} = \int dt w(t) P_{\alpha,t}
\een
All three recovery maps, $P_\alpha, P_{\alpha,t}, R_\alpha$ reduce to Bayes' inversion formula in the classical case. 
It turns out however that $R$ for the specific weight $w(t) = \frac{ \pi}{\cosh(2\pi t ) +1}$ understood from now on 
is a recovery channel with the desired property. Namely, 
one can show 
\ben
\label{iDPI}
S(\rho|\sigma) - S(\tilde \alpha(\rho)|\tilde \alpha(\sigma)) \ge -\log F(\rho|\tilde R_\alpha \tilde \alpha(\rho)),
\een
where $F(\omega_1 |\omega_2) = ( \tr\sqrt{\sqrt{\omega_1} \, \omega_2^{} \sqrt{\omega_1} } )^2$ is the fidelity between two density matrices, 
which has a well-defined generalization to state functionals on arbitrary von Neumann algebra types where density matrices may not exist. 
For general von Neumann algebras, these statements were first shown in \cite{hollands1} (for inclusion channels), and for type I in \cite{junge}.  
If you don't like the fidelity you can replace the right side also by $(1/4) \|\rho - \tilde R_\alpha \tilde \alpha (\rho)\|^2_{\rm Tr}$ where the norm between the states 
is the ``trace norm'' (defined also for general von Neumann algebras), or by the measured relative entropy $S_M(\rho | \tilde R_\alpha \tilde \alpha (\rho))$. 
For proofs of these more general statements and related material, see \cite{recovery1,recovery2,recovery3}. 

\medskip
\noindent
{\bf Exercise 17.} Show that if $\omega_i = |\chi_i\rangle\langle \chi_i|$ are pure states on the algebra of all bounded operators, then their fildelity is $|\langle \chi_1 | \chi_2\rangle |^2$, i.e. the 
transition amplitude. Show that for any pair of density matrices $F \le 1$ and $F=1$ if and only if $\omega_1 = \omega_2$.

\medskip
\noindent
The inequality \eqref{iDPI} is called an ``improved'' DPI. Its proof goes beyond these lecture notes but I mention that the operator $V$ used in 
the proof of the ordinary DPI also plays a major role, together with a host of other methods such as non-commutative $L_p$ norms, methods from complex analysis, and 
an interpolation theory for these norms. 

Let us instead get a feeling for what the improved DPI says. Suppose first that $S(\rho|\sigma) - S(\tilde \alpha \rho|\tilde \alpha \sigma) = 0$ exactly. 
Then $F(\rho|\tilde R_\alpha \tilde \alpha (\rho)) = 1$ and hence $\rho=\tilde R_\alpha \tilde \alpha (\rho)$ by exercise 17, so the state $\rho$ is recovered from $\tilde \alpha(\rho)$
exactly. 

\section{DPI in the sky: Generalized second law}

With the help of the ordinary DPI and the idea of half sided modular inclusion one can obtain 
the ``generalized second law'' of black hole thermodynamics in Schwarzschild spacetime in 
a semi-classical setting. The idea to use quantum information theoretic inequalities to 
get such a law first appeared in \cite{sorkin} to my knowledge. That the relative entropy should play 
a role in this context was observed by \cite{casini}. The reasoning that we shall present is essentially due to \cite{wall1,wall2,wall3} though I will 
emphasize more strongly the importance of half-sided modular inclusions; see also 
\cite{witten2} for a similar discussion.

Let us consider the exterior of the black hole and  a thermal equilibrium state with respect to the time translations. I call the exterior the ``right wedge'' $W$
by analogy with Rindler space. It can be shown that this state can be extended to the analytic extension of the Schwarzschild spacetime, if and only if the inverse temperature is 
$\beta_H=2\pi/a$, i.e. the inverse Hawking temperature (here $a=1/(2M)$ is the surface gravity, often denoted by $\kappa$). This state $|\Omega\rangle=|{\rm HH}\rangle$ is called the ``Hartle-Hawking (HH) state''. Informally speaking, the HH state is characterized by the fact that its reduced density matrix to $W$ is 
\ben
\omega_{\rm HH} = {\rm Tr}_{W'}  |{\rm HH}\rangle \langle {\rm HH} | = e^{-\beta_H K}
\een
where $K$ is the generator of time translations in $W$, i.e. up to an infinite additive constant (due to the sharp cutoff of the integral at the bifurcation surface where $H^+$ ends)
\ben
\label{log0}
K = \int_{H^+} \Theta_{\mu\nu} t^\mu dS^\nu + \int_{I^+} \Theta_{\mu\nu} t^\mu dS^\nu 
\een
Here $t^\mu$ is the time translation vector field, given by $t^\mu = \delta^\mu_t$ in the standard coordinates $(t,r,\theta,\varphi)$. 
On $H^+$, it is given by $t^\mu=ax^+ (\partial_+)^\mu$ where $x^+$ is an affine parameter such that $x^+=0$ on the bifurcation surface where 
$H^+$ and $H^-$ meet each other\footnote{Note that such an affine parameter is not unique; we may redefine $x^+ \to T(z^A)x^+$, where 
$z^A = (\theta,\varphi)$ and $T(z^A)$ is an arbitrary positive (smooth) function.}. In other words, in terms of this affine parameter
it is basically a boost, and you should note the close analogy between these formulas and \eqref{remat} in Rindler space (wedge). 

Of course, the reduced density matrix is only a formal object; what is really defined is the modular operator for the algebra $\cA_W$, related by 
``$\Delta_{\rm HH} = \omega_{\rm HH}^{} \otimes \omega_{\rm HH}^{-1}$''. 
Thus, 
\ben
\label{log0}
\log \Delta_{\rm HH} = -\beta_H \int_{H^+} \Theta_{\mu\nu} K^\mu dS^\nu - \beta_H \int_{I^+} \Theta_{\mu\nu} K^\mu dS^\nu + \text{left wedge}
\een
where ``left wedge'' is a corresponding contribution from the left (opposite) wedge not drawn in the figure which is over 
$H^-$ and $I^-$ in that wedge. 

Consider next a Cauchy surface $\Sigma_T$ as in the following figure.
\begin{figure}[h!]
\centering
\begin{tikzpicture}[scale=1.2]
\draw[snake] (0,2) -- (2,2);
\draw[snake] (0,-2) -- (2,-2);
\filldraw[color=black, fill=white, thin] (0,0) -- (2,2) -- (4,0) -- (2,-2) -- (0,0);
\filldraw[color=black, fill=lightgray, thin] (.5,.5) -- (2,2) -- (4,0) -- (2.5,-1.5) -- (.5,.5);
\filldraw[color=black, fill=gray, thin] (1,1) -- (2,2) -- (4,0) -- (3,-1) -- (1,1);
\draw[ thick] (1,1) -- (4,0);
\draw[thick] (.5,.5) -- (4,0);
\draw (2,.9) node[anchor=west]{$\Sigma_{T'}$};
\draw (2,0) node[anchor=west]{$\Sigma_T$};
\draw (.85,.85) node[anchor=east]{$H^+$};
\draw (3.85,.85) node[anchor=east]{$I^+$};
\end{tikzpicture}
\end{figure}
I can assume that with a suitable choice of affine parameter $x^+$ on the future horizon, this Cauchy surface intersects the horizon at $x^+=T$.
The domain of dependence of $\Sigma_T$ is called $W_T$. We again have the situation of a half-sided modular inclusion because points in $W_T$
do not leave $W_T$ for positive flow times of the time-translation symmetry: These would move $W_T$ to a region of the form $W_{T'}$ with Cauchy surface
$\Sigma_{T'}$ as indicated in the figure. 

Consider the reduced density matrix for this region, 
\ben
\label{logT}
\omega_{{\rm HH},T} = {\rm Tr}_{W'_T}  |{\rm HH}\rangle \langle {\rm HH} | = e^{-\beta_H K_T}
\een
By analogy with Rindler space \eqref{locboost}, we have
\ben
K_T = a \int\limits_{\{x^+>T\} \cap H^+} (x^+-T) \Theta_{\mu\nu} k^\mu dS^\nu + \int\limits_{I^+} \Theta_{\mu\nu} t^\mu dS^\nu
\een
where $k^\mu = (\partial_+)^\mu$ is the generator of affine translations $x^+ \to x^+ + const.$ on $H^+$, so $t^\mu = ax^+ (\partial_+)^\mu$ on $H^+$.
A different, less formal way of stating the same thing is to say that modular operator $\Delta_{{\rm HH},T}$ for the algebra 
$\cA(W_T)$ satisfies
\ben
\label{log}
\log \Delta_{\rm HH} - \log \Delta_{{\rm HH},T} = {\mathcal E}_{T,{\rm HH}} = aT \int_{H^+ \cup \bar H^-} \Theta_{\mu\nu} k^\mu dS^\nu,
\een
where we again encounter the ANEC operator on the right side.
Let us now consider another state $|\Psi\rangle$ and set $S_{\rm rel}[T] = S(\Psi | \Omega)_{W_T}$. 
The DPI, i.e. monotonicity of the relative entropy tells us again that 
\ben
S_{\rm rel}'[T] \le 0. 
\een
Going from the relative entropy to the entanglement entropy similar as in our discussion of the QNEC, I formally get using 
\eqref{log0}, \eqref{logT} that
\ben
S'_{\rm EE}[T] \ge  -2\pi   \int\limits_{H^+ \cap \{x^+>T\}} \langle \Theta_{\mu\nu} \rangle_{\Psi} k^\mu dS^\nu.
\een
Note that since $dS^\mu = k^\mu dx^+ dA$ where $dA=r_0^2 d^2 z$ is the area element on cross sections of $H^+$ parameterized by $z^A = (\theta,\varphi)$, 
the integrand on the right side the $++$ component of the stress tensor. 
In order to treat that term, we can appeal to the Raychaudhuri equation and assume the semi-classical Einstein equation \eqref{sclass} to relate 
$R_{++}$ to $ \langle \Theta_{++} \rangle_\Psi$:
\ben
\label{Raych}
\partial_+ \theta_+ = -\frac{1}{d-2} \theta_+^2 - \sigma_+^2 - R_{++} = -\frac{1}{2} \theta_+^2 - \sigma_+^2 - 8\pi G \langle \Theta_{++} \rangle_\Psi 
\een
The shear-squared $\sigma_+^2$ and the expansion-squared
$\theta_+^2$ are considered to be of higher order. Thus, under these assumptions/approximations we can write
\ben
 - \!\!\! \int\limits_{H^+ \cap \{x^+>T\}}  \langle \Theta_{++} \rangle_{\Psi} \, dA dx^+   \approx \frac{1}{8\pi G} 
 \int\limits_{H^+ \cap \{x^+>T\}} \partial_+ \theta_+ dA  dx^+.
\een
Now suppose $A[T]$ is the area of the horizon cross section at $x^+=T$. The expansion is the infinitesimal change in area cross section along a congruence of 
affine geodesics, so 
\ben
 \int\limits_{H^+ \cap \{x^+>T\}}  \partial_+ \theta_+ dA dx^+= -\int\limits_{H^+ \cap \{x^+=T\}}   \theta_+ dA =  - A'[T]
\een
Thus we see that
\ben
\partial_T\left( S_{\rm EE}[T] + \frac{A[T]}{4G} \right)  \equiv \partial_T S_{\rm gen} \gtrsim 0
\een
which is called the generalized second law. Note that the QFT piece should include all quantized perturbations around the spacetime background, including 
quantized gravitational perturbations. Note also that this derivation assumed a small change of the state $\Omega$ since we ignored the 
expansion and shear squared terms.
The  $\gtrsim$ should thus become exact for an infinitesimal change.

Perhaps it is worth mentioning that this derivation is independent of any of the objections raised in connection with the 
{\it Bekenstein bound}, see e.g., \cite{marolf1,marolf2,marolf3,marolf4,marolf5, marolf6}.
A valid criticism of the generalized second law as derived is that it is unnatural to consider black hole horizons as fundamental in a dynamical setting: The horizon
is a teleological concept being determined backwards in time once the entire evolution of the spacetime has been completed. Closer to the viewpoint of local physics 
is the concept of ``apparent horizon''. Unlike the concept of event horizon, it has the drawback that it is dependent on some conventions such as a slicing of 
spacetime into time-slices. Fortunately, to first order in the deviations from a stationary (non-dynamical) black hole with bifurcate Killing horizon, this non-uniqueness disappears. 
Since the above derivation is valid only for small deviations from a stationary black hole, this is in principle good enough. A version of generalized second law, involving a correspondingly modified notion of dynamical entropy, has been given by \cite{zhang}. Interestingly, its proof is not based on monotonicity of the relative entropy, but on 
the monotonicity of its first derivative (QNEC).

\section{Applications of improved DPI: QNEC, generalized second law, holographic reconstruction}

Let us apply the {\it improved} DPI to the situation where system $A$ is the wedge algebra $\cA_W$ in Minkowski spacetime, 
system $B$ is the shifted wedge algebra $\cA_{W_T}$, the channel 
$I: \cA_{W_T} \subset \cA_W$ is simply the inclusion. the dual channel $\tilde I$ on states is a ``partial trace'' over the causal complement of $W_T$ inside $W$, 
$\tilde I(\rho) = {\rm Tr}_{W_T' \cap W} \rho$. 
The reference state $\sigma_A$ is the vacuum state $|\Omega\rangle$ or rather its reduction to $W$, namely 
\ben
\sigma_A = {\rm Tr}_{W'} |\Omega\rangle\langle \Omega |.
\een
From now, I use the notation $S_{\rm rel}[T] = S(\Psi|\Omega)_{W_T}$ as e.g., in the discussion of the QNEC.
The improved DPI, or rather, a stronger version of it with the ``integral outside the fidelity'', can be shown to give \cite{hollands1} (we assume 
$T(z^A)=T$ to be a positive constant): 
\ben
\frac{1}{T} (S_{\rm rel}[0] -  S_{\rm rel}[T])\geq
 - \int_{T}^{\infty} \ln F_W( \Psi|e^{ix_+ P_+} \Psi)  \, \frac{dx_+}{x^2_+} .
\een
Here $F_W$ is the fidelity 
between the reduced density matrices of $|\Psi\rangle, e^{ix_+P_+}|\Psi\rangle$ to $W$.
$P_+$ is the momentum operator in the $x^+$ direction.

The above inequality can be seen as a cousin of the QNEC for finite $T>0$. To get a feeling for what it says, suppose that 
$\frac{1}{T} (S_{\rm rel}[0] -  S_{\rm rel}[T]) \approx 0$, i.e. a nearly zero rate of change of $S_{\rm rel}[T]$. For small $T$, the integral on the right side of the 
above inequality is dominated by the contribution from $x_+ \approx T$. 
Therefore $\ln F_W( \Psi|e^{iT P_+} \Psi) \approx 0$ and hence $F_W( \Psi|e^{iT P_+} \Psi) \approx 1$, 
so the reduced density matrices of $|\Psi\rangle$ and $e^{iT P_+}| \Psi \rangle$ to $W$ are nearly the same. This will be the case  
if $|\Psi\rangle$ is nearly an eigenstate for $e^{iT P_+}$ i.e. $|\Psi\rangle = \int dp_+ \psi(p_+)|p_+\rangle$ for some wave function $\psi(p_+)$
that is sharply peaked at integer multiples of $2\pi/T$. Intuitively, the position-space wave function $\hat \psi(x^+)$ of such a nearly  eigenstate is approximately 
periodic in $x^+$ with period $T$ as illustrated by the periodic cat images below.
 The inclusion channel basically discards the part of this wave function between $0 \le x^+ \le T$. However, for a nearly periodic 
wave function discarding any finite portion does not constitute a major loss of information, see the cat figure below, 
i.e. we can more or less restore the missing part by periodicity. In other words, it is 
intuitively clear that we can 
recover the original wave function with high precision if $\frac{1}{T} (S_{\rm rel}[0] -  S_{\rm rel}[T]) \approx 0$, just as the improved DPI says.

\begin{figure}[h!]
\begin{center}
  \includegraphics[width=0.9\textwidth,]{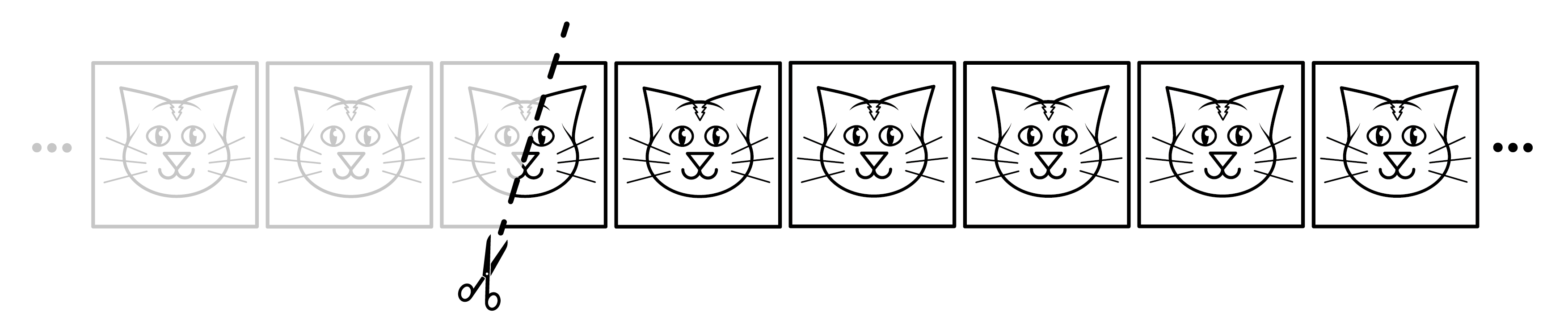}
  \end{center}
  \label{fig:18}
  \caption{Discarding part of a (nearly) periodic state does not constitute an essential loss of information because the part that has been chopped off can be restored.
  Here $T$ from the main text is analogous to the width of an individual cat image.}
\end{figure}

Since the improved DPI gives a better bound for any inequality derived from the ordinary DPI, we get an improved version of the generalized second 
law. Setting $S_{\rm gen} = S_{\rm bh} + S_{\rm qft}$ and denoting as above the partial states of $|\Psi\rangle$ on the Cauchy surfaces $\Sigma_0$ and $\Sigma_1$
by $\rho_0$ and $\rho_1$, we get for example
\ben
d S_{\rm gen} \gtrsim -\log F_{\Sigma_0}(\rho_0|\tilde R\rho_1).
\een
The recovery map $\tilde R$ is basically given by the same formula as in the case of Rindler horizons.
 As is intuitively clear, the increase in the generalized entropy controls the recoverability of the initial 
state on $\Sigma_0$ from the final state on $\Sigma_1$.

\medskip
Holographic ideas, especially in the context of the AdS-CFT correspondence, suggest a relation between entropies in the gravitational bulk and the boundary CFT, at least in a 
semi-classical regime. Such ideas are intimately related to the Ryu-Takaynagi proposal \cite{RT} and its generalization \cite{HR} for holographic calculations of the entanglement entropy. 
In particular, in \cite{Jafferis}, it was shown that bulk- and boundary relative entropies $S(\rho_a|\sigma_a)$ and $S(\rho_A|\sigma_A)$ 
similar to those discussed above are approximately equal (up to order $1/N_c$) under an appropriate 
identification of the bulk and boundary regions $a$ and $A$. More precisely $a$ is a certain region whose boundary inside AdS-spacetime (the ``HRRT surface'' \cite{RT,HR}) is 
cohomlogous to a boundary region $A$. There is a channel $N: \cA_A \to \cA_a$ -- the ``AdS-CFT correspondence'' -- mapping boundary to bulk operators (the construction of which 
goes through a certain procedure involving a ``code subspace'' in the boundary Hilbert space from an information theoretic viewpoint on AdS-CFT) and $\sigma_A = \tilde N(\sigma_a)$.  
The approximate equality between $S(\rho_a|\sigma_a)$ and $S(\rho_A|\sigma_A)$ yields a close connection to ideas of state recovery as described above, i.e. to a viewpoint 
where the AdS-CFT correspondence is seen as a kind of recovery problem: The inversion problem for the channel $N: \cA_A \to \cA_a$  is solved 
approximately by the twirled Petz map $R$, for many more details on such ideas see e.g., \cite{entanglement1,entanglement2,entanglement3,entanglement4}, which build on the ideas by \cite{Kabatetal}. For connections 
to the information loss paradox see e.g., \cite{pennington1,pennington2}. 

\medskip
\noindent
{\bf Acknowledgements:} I thank Tom Faulkner and Robert M. Wald for discussions about the generalized second law, Nima Lashkari for discussions about state recovery, 
and Thomas Endler from MPI-MiS for help with figures. I am 
grateful to the Max-Planck Society for supporting the collaboration between MPI-MiS and Leipzig U., grant Proj. Bez. M.FE.A.MATN0003.

\end{document}